\documentclass[]{interact}

\usepackage{epstopdf}
\usepackage{subfigure}

\usepackage{natbib}
\usepackage{float}

\DeclareMathOperator{\arccosh}{arccosh}
\DeclareMathOperator\erf{Erf}

\theoremstyle{plain}

\theoremstyle{definition}

\theoremstyle{remark}

\begin{document}


\title{Scalar-field potential for viable models in $f(R)$ theory}

\author{
\name{R.~A. Hurtado\textsuperscript{a}\thanks{CONTACT R.~A. Hurtado. Email: rahurtadom@unal.edu.co} and R. Arenas\textsuperscript{a}}
\affil{\textsuperscript{a}Observatorio Astronómico Nacional, Universidad Nacional de Colombia, Bogotá, Colombia.}
}

\maketitle

\begin{abstract}
The $f(R)$ theory of gravity can be expressed as a scalar tensor theory with a scalar degree of freedom $\phi$. By a conformal transformation, the action and its Gibbons-York-Hawking boundary term are written in the Einstein frame and the field equations are found. An effective potential is defined from part of the trace of the field equations in such a way that it can be calculated as an integral of a purely geometric term. This potential as well as the scalar potential are found, plotted and analyzed for some viable models of $f(R)$ and for two other proposed new, shown viable, models. 
\end{abstract}


\section{Introduction}
Just like GR, $f(R)$ gravity describes a dynamical tensorial field $g_{\mu\nu}$ that depends on the distribution of matter and energy and relates it to the geometry of the space-time through the Field equations, which are found as the critical points of an action written from a lagrangian density.
\\
The constant efforts to modify the General Relativity in order to apply it at large scales started about 1933, when F. Zwicky concluded from observational data that the dynamical behaviour of clusters of galaxies did not correspond to the theoretical predictions, deducing that the content of non-luminous matter of the Universe must be greater than the content of the baryonic matter [\cite{1933}]. This was the first hypothesis of existence of Dark matter (DM). Likewise, observational cosmology [\cite{Smoot1992td, Liddle:1992wi, Riess:1998cb, Perlmutter:1998np, Spergel:2006hy, Komatsu:2010fb}] has indicated that the Universe has experimented two phases of cosmic acceleration: Inflation [\cite{Guth:1980zm,Lyth:1998xn,Liddle:2000cg}] (before the radiation dominated epoch and required to solve the flatness and horizon problems, and needed to explains part of the spectrum of temperature anisotropies observed in the Cosmic Microwave Background [CMB]) and Dark Energy (DE)  [\cite{Straumann:2007zz,Copeland:2006wr,Frieman:2008sn}](after matter dominated epoch) or current acceleration expansion of the Universe. However, due to the attractive gravitational effects of the baryonic matter, it is impossible to explain these acceleration phases only from its positively contributions of pressure and energy density to the dynamic of the Universe. This is an open problem and these phases of acceleration are not fully understood.
\\
Among the theories trying to solve this problem, the $\Lambda$-Cold Dark Matter ($\Lambda$CDM) model [\cite{Carroll:2004st,Weinberg:1988cp,Peebles:2002gy,Riess:1998cb}] results when interpreting the Cosmological constant $\Lambda$ in the Einstein's field equations, as a new component of energy with negative pressure, together with DM model, however at this time, almost 90 years after the publication of Zwicky, the nature of the constituents of the DM has not been fully explained or detected experimentally, and there are a major problems to test the Dark Energy model, because the vacuum energy density strongly depends on the high-scale physics scenario, so that the required corrections are perturbatively unstable [\cite{Marsh:2016ynw,Wondrak:2017eao,Weinberg:1988cp}]. Furthermore, $\Lambda$CDM does not explain inflation epoch because the radiation dominated phase started when inflation ended.
\\
Alternatively, some models of Modified Gravity as the Scalar Tensor Gravity has been able to describe the galaxy rotation curves without the dark matter component [\cite{Moffat:2013sja,Moffat:2013uaa,Moffat:2014pia}]; and recent progress has been made in the $f(R)$ theory to explain the accelerated  expansion of the Universe without the necessity to introduce the cosmological constant [\cite{Song:2006ej,Tsujikawa:2010zza,Takahashi:2015ifa}]. The power of this theory relies on a higher order curvature gravity, and therefore it is expected to reproduce the Einstein Field equations with Cosmological constant ($\Lambda$CDM) when $f(R)=R-2\Lambda$. Thus $f(R)$ theory allows, for example, to generalize the functional form for the $\Lambda$CDM model, to $f(R)=a_1R+a_2R^2$, with $a_1$, $a_2$ constants. The effects of expansion traced in the term $a_2R^2$, have been studied first by Starobinsky, which constituted the first model for inflation [\cite{Starobinsky:1980te}]; and, as indicated by De Felice and Tsujikawa [\cite{DeFelice:2010aj}], this model is a very good alternative to scalar fields because it is well consistent with the temperature anisotropies observed in CMB.
\\
In spite of the progress in the modified gravity, it is not a widely accepted theory that satisfies the gravity constrictions [\cite{Amendola:2006we,Song:2007da,Thongkool:2009vf}] of the cosmological purposes at large structure as well as to the scales of Black Holes, making a complicated work to find a $f(R)$ model [\cite{Saidov:2010wx,Mijic:1986iv,Erickcek:2006vf,Tsujikawa:2007xu,Baghram:2007df,Kluson:2009xx}].

This paper is presented as follows: in section \ref{sec:fieldequations}, action and field equations in $f(R)$ theory are described, in section \ref{sec:scalartensor} the equivalence between $f(R)$ gravity and scalar tensor theories is analyzed and the effective potential is defined, while in section \ref{sec:viablemodels} the potential of some viable models are studied and plotted. The conclusions of the work are given in section \ref{sec:discussion}.

\section{Field equations}\label{sec:fieldequations}
Field equations in $f(R)$ gravity are found from an action that generalizes the Einstein-Hilber (E-H) action, $\int d^4x\sqrt{-g}R$, expressed as an arbitrary function of the scalar curvature, that is [\cite{Sotiriou:2008rp,DeFelice:2010aj})
\begin{equation}\label{actionf}
    I_f=\int_{\Sigma}d^4x \sqrt{-g}f(R)
\end{equation}
defined over a hypervolume $\Sigma$, plus a Gibbons-York-Hawking boundary ($\partial\Sigma$) term [\cite{Dyer:2008hb,Guarnizo:2010xr}]
\begin{equation}
    I_{m}=I_{f}+I_{GYH},
\end{equation}
where the subscript $m$ refers to modified action, and 
\begin{equation}
    I_{GYH}=2\int_{\partial\Sigma}d^{3}x\varepsilon\sqrt{|h|}n^{\mu}_{;\mu}F(R),
\end{equation}
and $F(R)=f'(R)=df(R)/dR$, with $h$ the determinant of the induced metric, defined as
\begin{equation}
    g^{\mu\nu}=h^{\mu\nu}+\varepsilon n^\mu n^\nu,
\end{equation}
and $n_\mu$ the unit normal to $\partial \Sigma$, and $\varepsilon=n_\mu n^\mu$ equal to 1 (-1) if $\partial\Sigma$ is timelike (spacelike) hypersurface. If a contribution of matter $I_M$ is also taken into account [\cite{Carroll:2004st}], the total action is then
\begin{equation}\label{totalaction}
    I=\frac{1}{2\kappa}I_{m}+I_{M}.
\end{equation}
Applying the principle of least action\footnote{This work only takes into account the Metric formalism of the $f(R)$ theory, in which Christoffel symbols are dependent on the metric tensor $g_{\mu\nu}$ and equations of motion are found varying the action with respect to the same.
} and noting that some terms are canceled at the boundary [\cite{Guarnizo:2010xr}], the field equations in the $f(R)$ theory are obtained as
\begin{equation}\label{fieldeq}
    F(R) R_{\mu\nu}-\frac{1}{2}g_{\mu\nu}f(R)-F(R)_{;\mu\nu}+g_{\mu\nu}F(R)_{;\alpha}^{;\alpha}=\kappa T_{\mu\nu},
\end{equation}
where $F(R)_{;\mu\nu}=\nabla_\mu\nabla_\nu F(R)$ and $g^{\alpha\beta}F(R)_{;\alpha\beta}=F(R)_{;\alpha}^{;\alpha}=\square F(R)$ is the D'Alembertian for the function $F(R)$. Field equations (\ref{fieldeq}) are a set of fourth order partial differential equations of the metric, and Einstein field equations are obtained when $f(R)=R$.

\section{Modified gravity as scalar tensor theory}\label{sec:scalartensor}
One of the most studied alternative theories to General Relativity is the Brans-Dicke theory [\cite{brans1961}], which is a scalar-tensor theory of gravity described by the scalar field $\phi$ responsible for mediate the gravitational interaction. There is a lot of written literature on this subject [\cite{campanelli,chau,romero,matsuda,hawkingbhbd}], and it is one of the few theories that remains valid since it has been able to overcome all the available observational tests [\cite{will2018}]. Brans-Dicke theory is defined from an action [\cite{Faraoni_1999,Kofinas_2016}] plus a boundary term
\begin{equation}\label{bransdickeaction}
    I_{BD}=\frac{1}{2\kappa}\int_{\Sigma} d^4x \sqrt{-g}\left[\phi R-\frac{\omega}{\phi}g^{\mu\nu}\phi_{;\mu}\phi_{;\nu}-V(\phi)\right]+\frac{1}{\kappa}\int_{\partial\Sigma}d^{3}x\varepsilon\sqrt{|h|}n^{\mu}_{;\mu}\phi+I_M,
\end{equation}
where $I_M$ does not depend on the scalar field $\phi$, the parameter $\omega$ is called the dimensionless Brans-Dicke coupling constant and $V(\phi)$ is the scalar-field potential. An important fact of $f(R)$ gravity is that its action can be written in the form of the Brans Dicke theory and in this way, it can be expressed as a scalar tensor theory [\cite{Faraoni:2010yi,kim,Sotiriou:2008rp}]. This can be easily shown if we write the action (\ref{totalaction}) as 
\begin{equation}\label{action5}
    I=\frac{1}{2\kappa}\int_{\Sigma}d^4x\sqrt{-g}\left\{F(R) R-[F(R) R-f(R)]\right\}+\frac{1}{\kappa}\int_{\partial\Sigma}d^{3}x\varepsilon\sqrt{|h|}F(R)n^{\mu}_{;\mu}+I_{M},
\end{equation}
so, in the case where the parameter $\omega$ vanishes, comparing the two actions it is immediately seen that
\begin{equation}\label{suppositions}
    \phi(\psi)=F(\psi), \quad\text{and} \quad V(\psi)=F(\psi) \psi-f(\psi),
\end{equation}
so, we think of $\phi$ as the degree of freedom scalar field, while the field $\psi$ is directly associated to the scalar curvature. In this way, action (\ref{actionf}) is rewritten as
\begin{equation}\label{actionjordan}
    I_{f}=\int_{\Sigma}d^4x\sqrt{-g}\left[\phi R-V(\psi)\right].
\end{equation}
Taking the variation of (\ref{bransdickeaction}) with respect to $g^{\mu\nu}$ and in presence of matter, field equations are found
\begin{equation}
    G_{\mu\nu}\phi+g_{\mu\nu}\phi_{;\alpha}^{;\alpha}-\phi_{;\mu\nu}+\frac{\omega}{\phi}\left(\frac{1}{2}g_{\mu\nu}\phi_{,\alpha}\phi^{,\alpha}-\phi_{,\mu}\phi_{,\nu}\right)+\frac{1}{2}g_{\mu\nu}V(\phi)=\kappa T_{\mu\nu},
\end{equation}
which reproduce the field equations (\ref{fieldeq}) when relations (\ref{suppositions}) are fulfilled and $\omega=0$.
Action (\ref{actionjordan}) is written in the so-called Jordan frame, nevertheless if we perform a conformal transformation of the metric $\tilde{g}_{\mu\nu}=F(\psi) g_{\mu\nu}=\phi(\psi) g_{\mu\nu}$, the scalar curvature is expressed as
\begin{equation}
    R=\phi(\psi)\left[\bar{R}+3\frac{\phi'(\psi )}{\phi(\psi )}\bar{\psi }_{;\mu }^{;\mu }+3\left(\frac{\phi''(\psi )}{ \phi(\psi)}-\frac{\phi'(\psi )^2}{2\phi(\psi )^2}\right)\psi_{,\mu }\bar{\psi }^{,\mu }\right],
\end{equation}
where $\bar{\psi}_{;\mu}^{;\mu}=\bar{g}^{\mu\nu}\psi_{;\mu \nu}$, $\bar{\psi}^{,\mu}=\bar{g}^{\mu\nu}\psi_{,\nu}$, and the action (\ref{action5}) is transformed as
\begin{multline}\label{bdactioninf}
    I=\frac{1}{2\kappa}\int_{\Sigma} d^4x\sqrt{-\bar{g}}\left[\bar{R}+3\left(\frac{\phi''(\psi )}{ \phi(\psi)}-\frac{\phi'(\psi )^2}{2\phi(\psi )^2}\right)\psi_{,\mu }\bar{\psi }^{,\mu }+3\left(\frac{\phi '(\psi )}{\phi (\psi )}\right)_{;\mu \nu }\bar{g}^{\mu \nu }\psi-\bar{V}(\psi)\right]+\\
    \frac{3}{2 \kappa }\int_{\partial\Sigma} d^3x\epsilon \sqrt{|\bar{h}|}\frac{1}{\phi(\psi)}\left\{\left[1+\psi  \left(\frac{\phi '(\psi )}{\phi (\psi )}-\frac{\phi ''(\psi )}{\phi '(\psi )}\right)\right]\phi '(\psi )\bar{\psi }^{,\mu }n_{\mu }+\frac{2}{3}n^{\mu}_{;\mu}\right\}+I_{M},
\end{multline}
and with the potential defined as
\begin{equation}\label{potentialvtilde}
    \bar{V}(\psi)=\frac{V(\psi)}{\phi(\psi)^2},
\end{equation}
action (\ref{bdactioninf}) it is said to be written in the Einstein frame of the theory. In particular, the transformation
\begin{equation}\label{transformation1}
    \phi(\psi)=c e^{\sqrt{\frac{2\kappa}{3}}\psi},
\end{equation}
with $c$ some constant, produces the action  
\begin{equation}
    I=\frac{1}{2\kappa}\int_{\Sigma} d^4x\sqrt{-\bar{g}}\left[\bar{R}+\kappa\psi_{,\mu }\bar{\psi }^{,\mu }-\bar{V}(\psi)\right]+\int_{\partial\Sigma} d^3x\epsilon \sqrt{|\bar{h}|}\left(\sqrt{\frac{3}{2\kappa}}\bar{\psi }^{,\mu }n_{\mu }+\frac{1}{\kappa}n^{\mu}_{;\mu}\right)+I_{M},
\end{equation}
in which the scalar field $\psi$ is coupled minimally with matter [\cite{Sotiriou:2008rp}].
\\
These fields redefinitions only affect the form and not the background structure of the action itself; however, how can we be sure that it is indeed the same theory with different representation? The answer could escape the scope of this work, nevertheless a necessary condition, but maybe not sufficient, is the fact that both theories must describe equivalently the dynamics of any system, which means that the field equations are the same from a mathematical point of view, and as can be seen, variation of action (\ref{bdactioninf}) produces
\begin{equation}
    \phi (\psi ) \bar{G}_{\mu\nu}+\frac{3 \phi '(\psi )^2}{2 \phi (\psi )}\left(\frac{1}{2}\psi _{,a}\bar{\psi }^{,a}\bar{g}_{\mu\nu}-\psi _{,m}\psi _{,n}\right)+\frac{1}{2} \bar{V} \phi (\psi ) \bar{g}_{\mu\nu}=\kappa  T_{\mu\nu},
\end{equation}
which are exactly the same as Eq. (\ref{fieldeq}) under the inverse transformation $g_{\mu\nu}=\phi^{-1}\bar g_{\mu\nu}$.

Once defined $f(\psi)$ the behaviour of $V(R)$ is determined by 
\begin{equation}
    \frac{d V(\psi)}{d\psi}=\phi(\psi),
\end{equation}
thus
\begin{equation}
    f(\psi)=\psi V'(\psi)-V(\psi),
\end{equation}
however, the evolution of the potential must be determined by the distribution of mass, whose dependence is provided by the trace equation
\begin{equation}\label{tracescalar}
    \phi \psi-2 f(\psi)=\kappa T-3\phi_{;\alpha }^{;\alpha},
\end{equation}
so
\begin{equation}
    \frac{d V(\phi)}{d\phi}-\frac{2 V(\phi)}{\phi}=\frac{1}{\phi}\left(3\phi_{;\alpha}^{;\alpha}-\kappa T\right),
\end{equation}
this is a first order differential equation whose left hand side can be expressed as a total derivative by multiplying both sides for the integrating factor $\phi^{-2}$, that is
\begin{equation}
    \frac{d}{d\phi}\left(\frac{V(\phi)}{\phi^2}\right)= \frac{1}{\phi^3}\left(3\phi_{;\alpha}^{;\alpha}-\kappa T\right),
\end{equation}
whose solution is
\begin{equation}
    V(\phi)=\phi^2\int\frac{1}{\phi^3}\left(3{\phi}_{;\alpha}^{;\alpha}-\kappa  T\right)d\phi,
\end{equation}
this integral depends on the matter distribution as well as the d'Alembertian of the function $\phi$, which in turn is function of $\psi$. In order to solve the integral, we define the effective potential $v=v(\phi)$ as
\begin{equation}
    \frac{d v}{d \phi}=3\phi_{;\alpha}^{;\alpha}-\kappa T,
\end{equation}
so
\begin{equation}\label{scalarpotential1}
    V(\phi)=\phi^2 \int\frac{1}{\phi^3}\frac{d v}{d\phi}d\phi,
\end{equation}
however, by the chain rule the derivative of the potential $v$ is expressed in terms of $\psi$,
\begin{equation}
    \frac{d v}{d \phi}=\frac{d v}{d\psi}\frac{d\psi}{d \phi},
\end{equation}
and Eq. (\ref{tracescalar}) gives the potential as an integral
\begin{equation}\label{potentialv}
    v(\phi)=\int \left[2 f(\psi)-\phi \psi\right]\phi'd\psi.
\end{equation}
In practice, given any $f(R)$ model, Eq. (\ref{potentialv}) is evaluated replacing the values of $R$ obtained from the equation $F-\phi=0$. However, let us looking for the particular case when $V(\phi)=v(\phi)$, which implies that
\begin{equation}
    \int \left[2 f(\psi)-\phi(\psi) \psi\right]\phi'(\psi)d\psi=\phi(\psi) \psi-f(\psi),
\end{equation}
or
\begin{equation}
    \left[2 f(\psi)-\phi(\psi) \psi-\psi\right]\phi'(\psi)=0.
\end{equation}
If $\phi'(\psi)\neq0$, and $R=\psi$, the solution is $f(\psi)=\psi+\mu \psi^2$, $\mu$ is constant, which constitutes the Starobinsky model proposed around 1980 [\cite{Starobinsky:1980te,Starobinsky:1983zz,Mukhanov:1981xt,Ferrara:2014ima,Asaka:2015vza,Paliathanasis:2017apr}], which describes inflation from the higher order gravitational term, $\psi^2=R^2$. For this model the scalar curvature increases linearly with respect to $\phi$, $\mu R=(\phi-1)/2$ and the effective potential as a function of $\phi$ is a parabola with axis at $\phi=-1$, that is $v(\phi)=\mu R^2=(\phi-1)^2/4\mu$.

In next section the potentials $V(\phi)$ and $v(\phi)$ will be found for some models of $f(R)$.
\section{Viable models}\label{sec:viablemodels}
It is known that any model of $f(R)$ gravity should satisfy the following observational tests (summarized in [\cite{Hu:2007nk}])
\begin{itemize}
    \item Cosmic Microwave Background: the theory must be in asymptotic correspondence with $\Lambda$-Cold-Dark-Matter ($\Lambda$CDM) model for high-redshift regime.
    \item Accelerated expansion of the Universe without a cosmological constant.
    \item Low redshift regime: constraints from the Solar system and the Equivalence Principle
\end{itemize}
Fortunately, there are several viable models satisfying these tests, some have been fairly studied as alternative models to Dark Matter within the framework of $f(R)$ theory. For example $1/R$, $R^n$ and $R^{-n}+R^{m}$ with $n,m>0$, describe accelerated expansion and early time inflation [\cite{Capozziello:2002rd,Nojiri:2006ri,Brookfield:2006mq,Li:2007xn,Rador:2007gq,Nojiri:2003ni,Cognola:2005de,Nojiri:2006su,Nojiri:2006je,Song:2006ej}]; the first model pass the Solar System test [\cite{Nojiri:2006ri,Nojiri:2003ft}], while in the case of $R+(-R^{\beta})$ it is constrained from the Wilkinson Microwave Anisotropy Probe (WMAP) data $\beta\sim10^{-3}$ and from Supernova Lagacy Survey (SNLS) and Sloan Digital SkySurvey (SDSS) data $\beta\sim10^{-6}$ [\cite{Li:2006ag}] and $\beta\sim3\cdot10^{-5}$ [\cite{Koivisto:2006ie}]; logarithmic models  $R+\ln(R/\alpha)$ constrained by the weak energy condition and the current values of the derivatives of the scale factor of the Friedmann-Robertson-Walker, giving $\alpha\sim1.2x10^{-41}m^{-2}$ [\cite{PerezBergliaffa:2006ni}]. New works based on parameterized functions, point towards the reconstruction of models from observations, either of evolution of the Universe or of the large scale structure [\cite{Lee:2017lud}].

In next sections we investigate the behaviour of the effective and scalar-field potentials for some viable models, taking into account the aforementioned observational tests and according to the following conditions derived in [\cite{Hu:2007nk}], rewriting $f(R)=R+\tilde f(R)$
\begin{equation}\label{condition1}
    \lim_{R\to \infty}\tilde f(R)=const,
\end{equation}
and
\begin{equation}\label{condition2}
    \lim_{R\to 0}\tilde f(R)=0.
\end{equation}
\subsection{Starobinsky model}
\begin{figure}[ht]
\centering
\subfigure[Starobinsky, Eq. (\ref{staro2007})]{
\resizebox*{8.5cm}{!}{\includegraphics{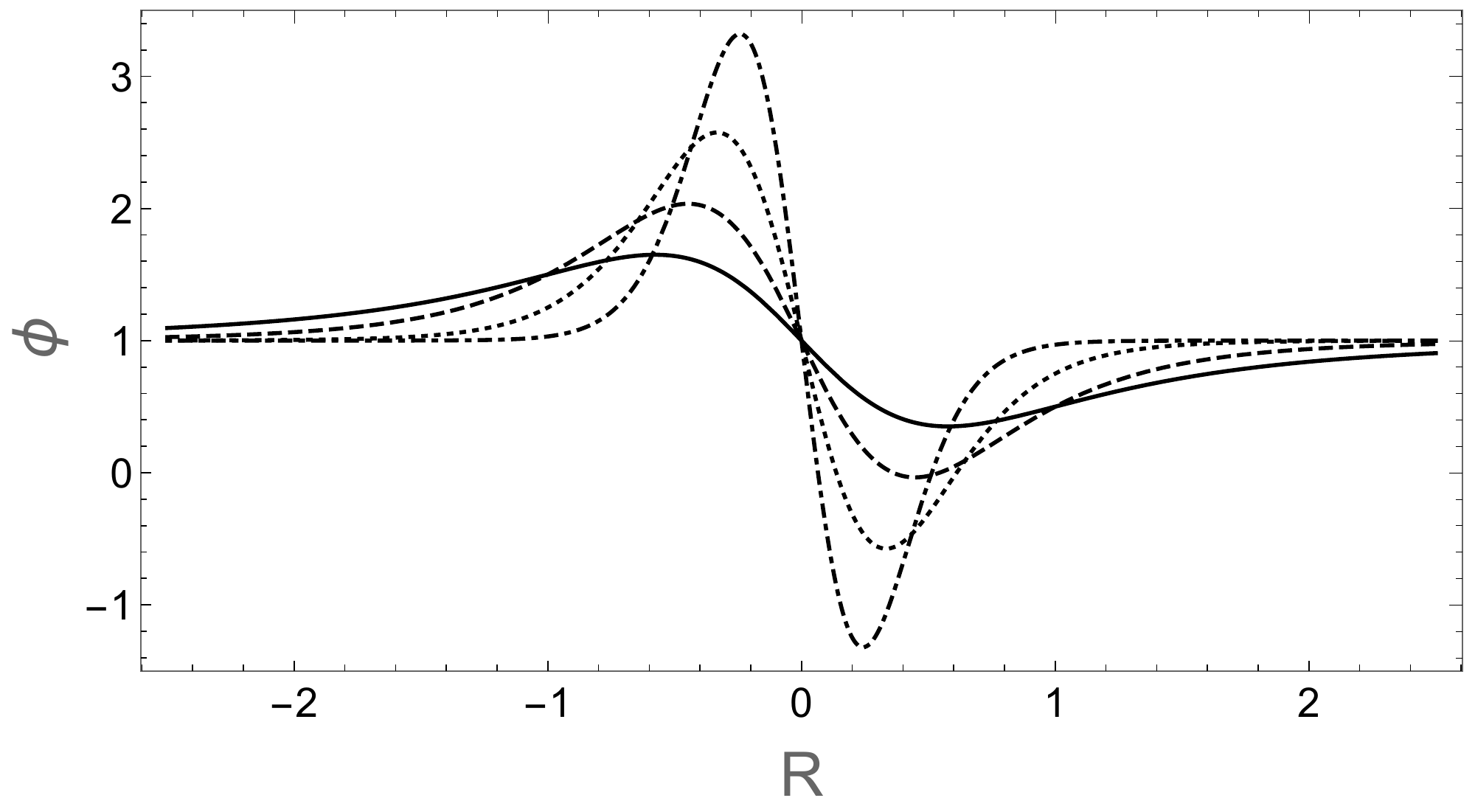}}}\hspace{5pt}
\subfigure[Hu-Sawicki, Eq. (\ref{husawicki})]{
\resizebox*{8.5cm}{!}{\includegraphics{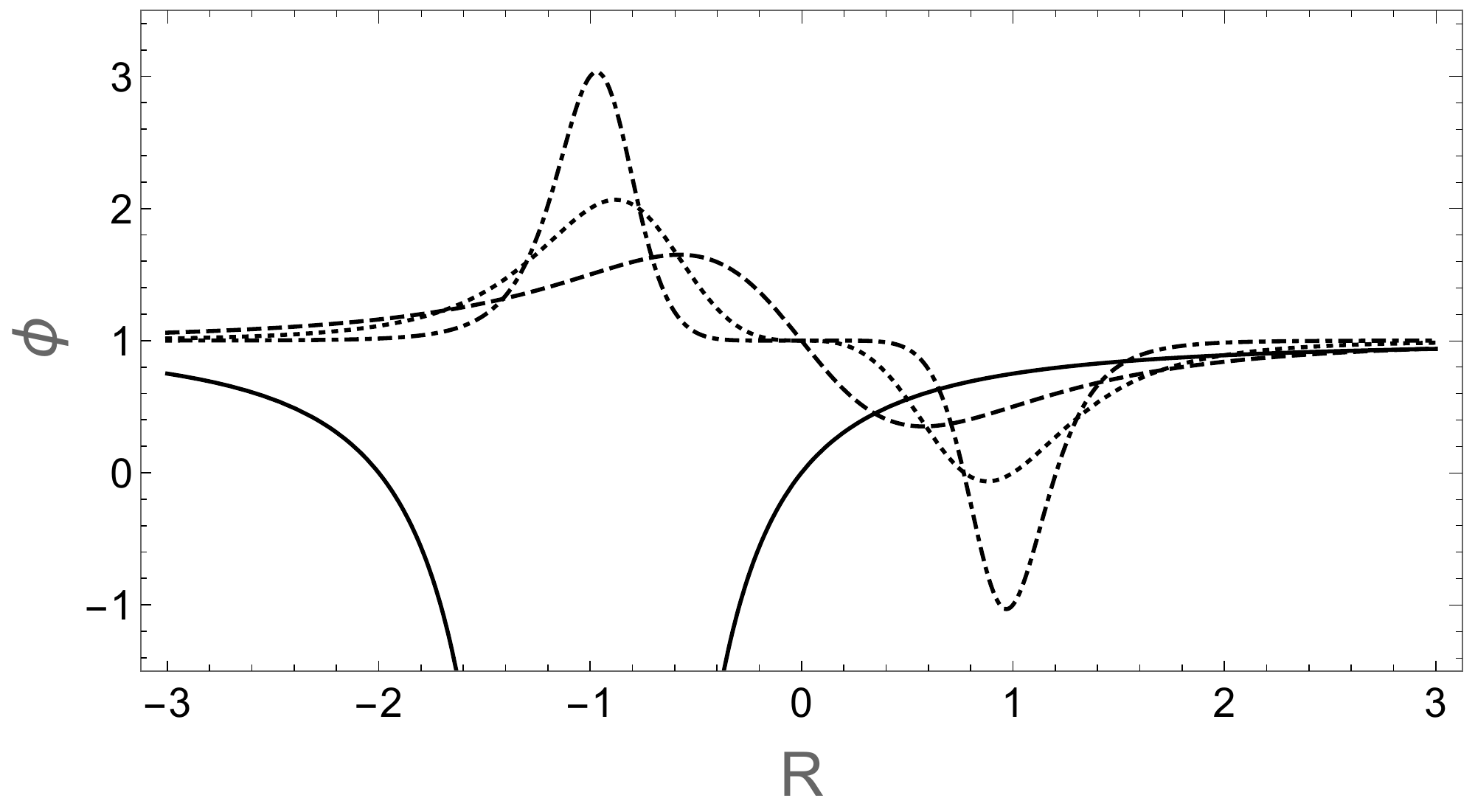}}}
\\
\subfigure[Complementary exponential model, Eq. (\ref{anotherexp})]{
\resizebox*{8.5cm}{!}{\includegraphics{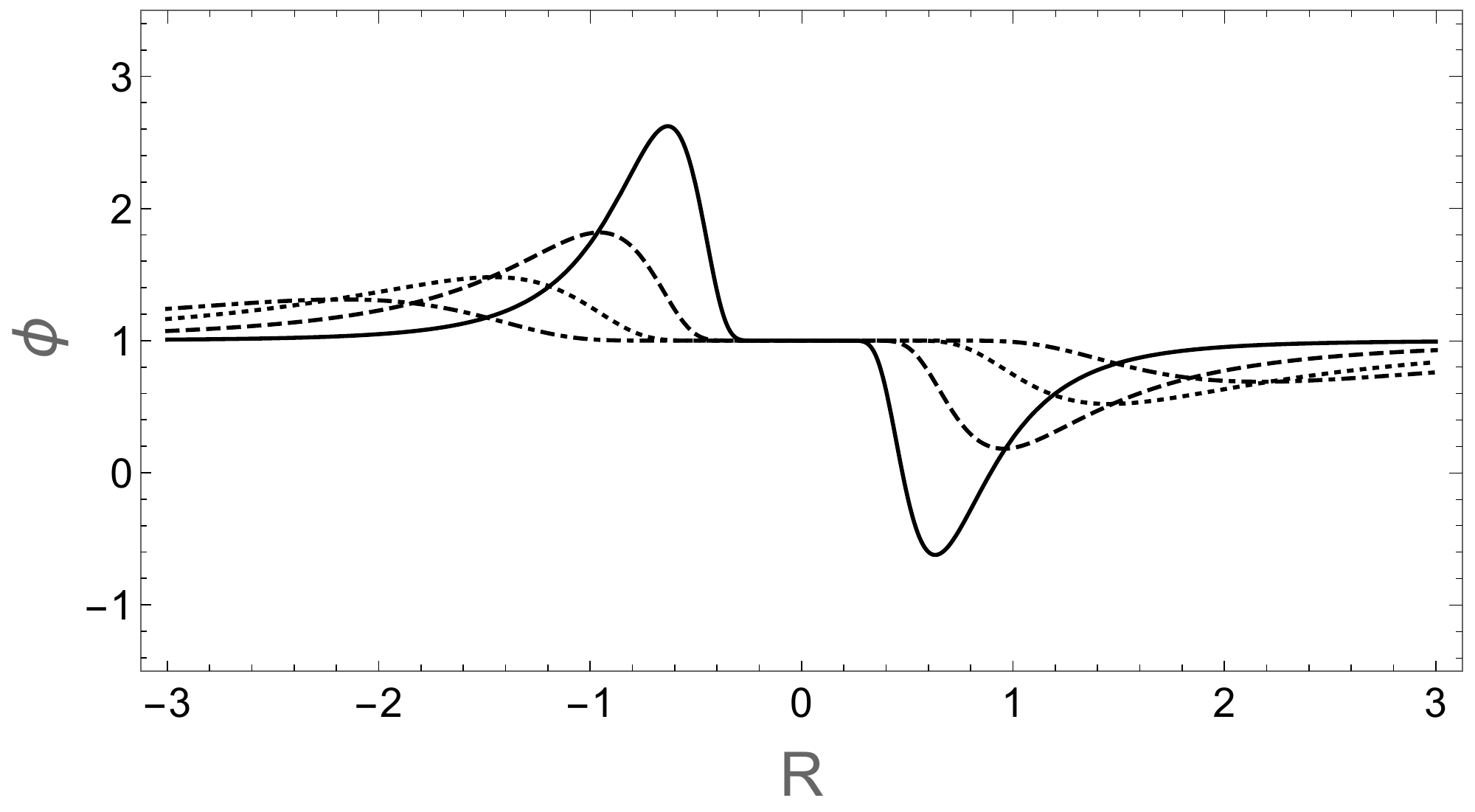}}}\hspace{5pt}
\subfigure[Modified Starobinsky model, Eq. (\ref{modifiedstarobinsky})]{
\resizebox*{8.5cm}{!}{\includegraphics{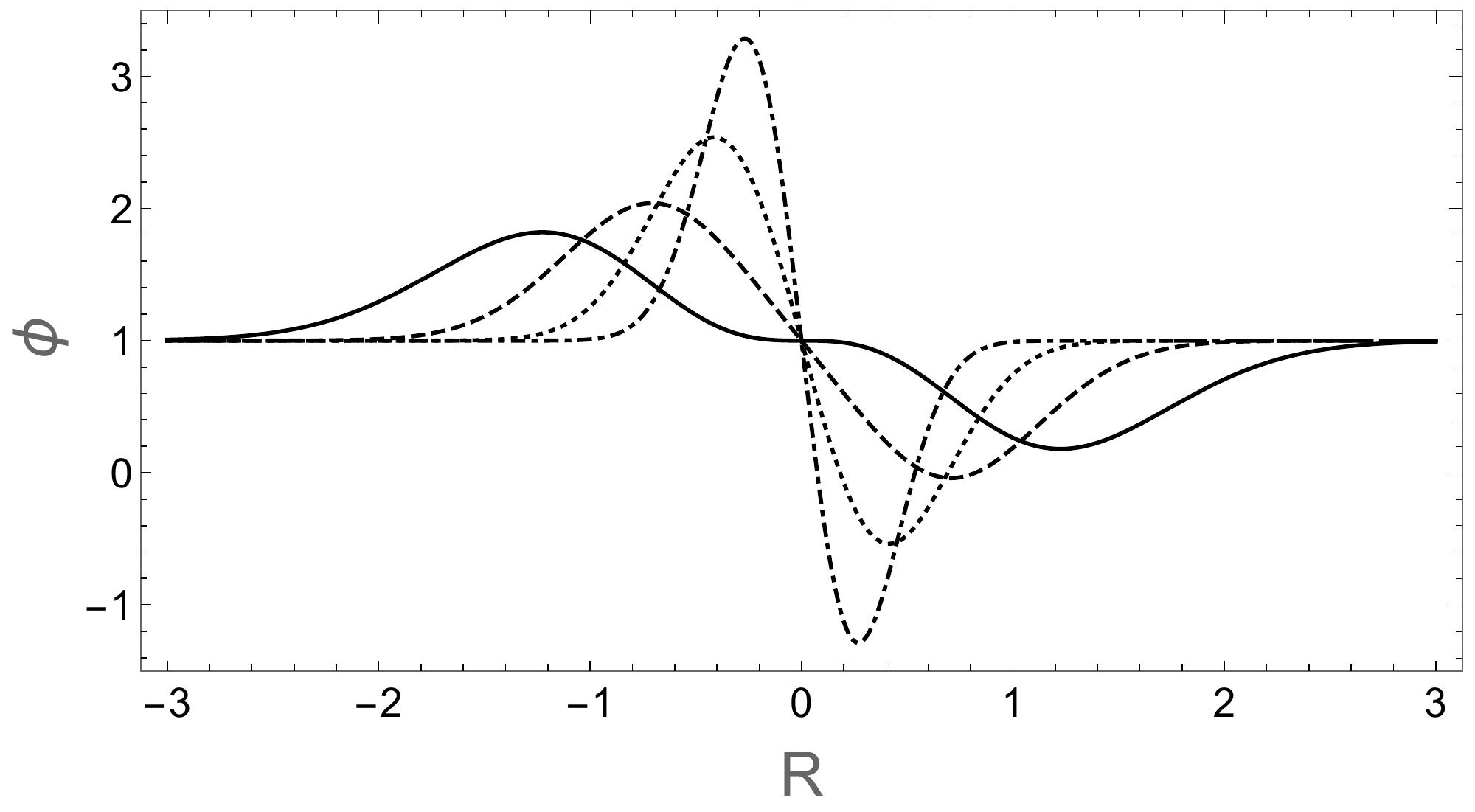}}}
\caption{Scalar field as a function of the scalar curvature for some values of power $n=1, 2, 4, 8$, continuous, dotted, dashed, dot-dashed lines, respectively. All models, except Hu-Sawicki have continuos solutions with  two maximum points and in (b), (c) and (d) there is at least one curve with zero slope transition region between those extremal points, this will have implications in the form of the effective and scalar potentials.} \label{fig:staro3}
\end{figure}

\begin{figure}[ht]
\centering
\subfigure[Starobinsky]{
\resizebox*{8.5cm}{!}{\includegraphics{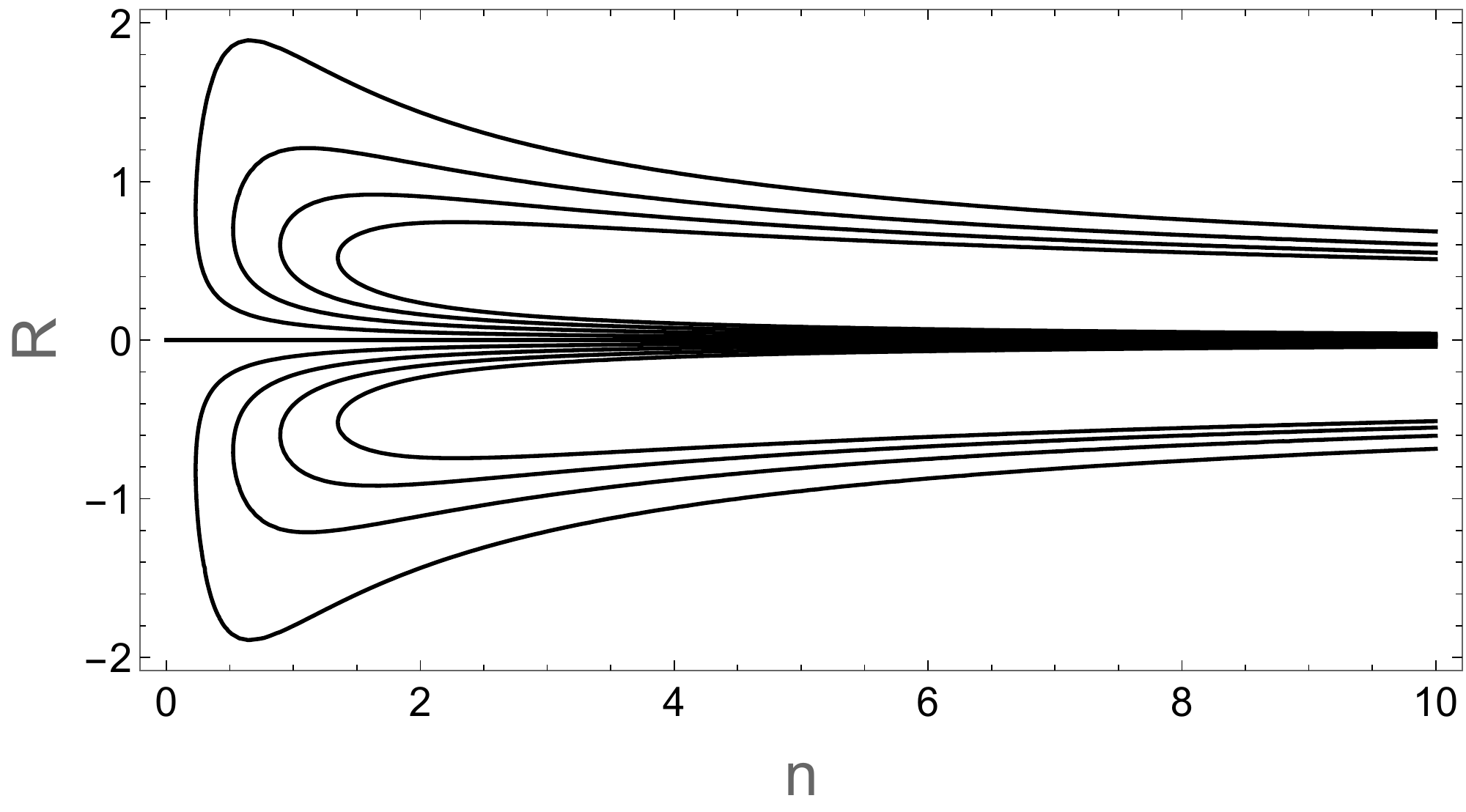}}}\hspace{5pt}
\subfigure[Hu-Sawicki]{
\resizebox*{8.5cm}{!}{\includegraphics{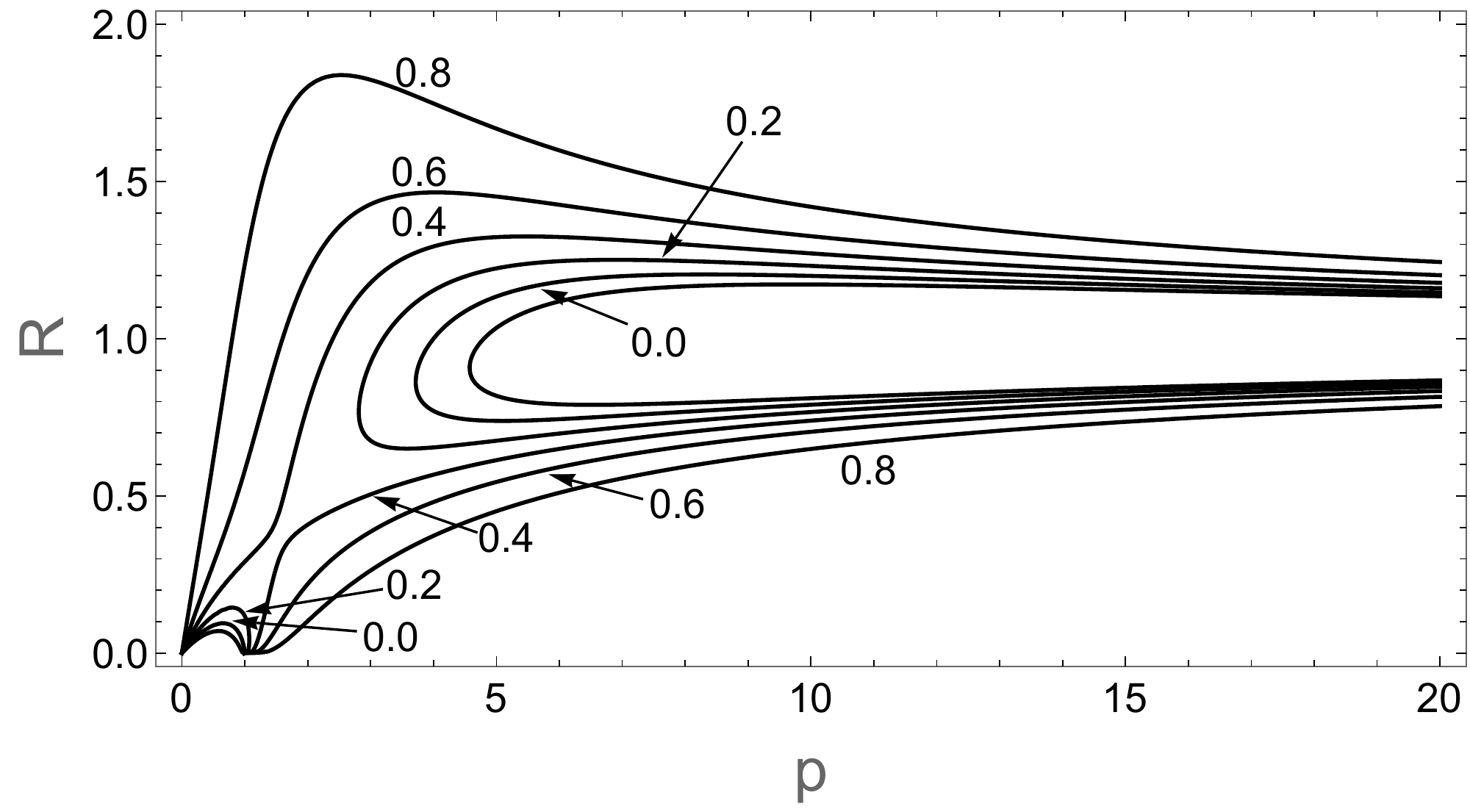}}}
\caption{Scalar curvature as a function of the parameter $n$. In (a) from bottom to top and left to right $\phi=$-1.2, -1.4, -1.6, -1.8, -0.8, -0.6, -0.4, -0.2, $R=0$ is obtained when $\phi=0$.} \label{staro1}
\end{figure}

Let us consider the model proposed by Starobinsky [\cite{Starobinsky:2007hu}] in order to satisfy cosmological and Solar system constraints
\begin{equation}\label{staro2007}
    f(R)=R+\mu R_0\left[\left(1+\frac{R^2}{R_0^2}\right)^{-n}-1\right],
\end{equation}
with $n,\mu>0$ parameters of the model, and $R_0$ is the constant characteristic curvature. The relation between $R$ and $\phi$
\begin{equation}\label{phireq}
    1-\phi-2 \mu  n x \left(1+x^2\right)^{-n-1}=0.
\end{equation}
with $x=R/R_0$, cannot be inverted analytically for any value of the parameters, however in order to simplify the model we fix $R_0=1$. The behaviour of $\phi$ with respect to $R$ is shown in Fig. \ref{fig:staro3} for some values of $n$ and taking $\mu=1$, and in Fig. \ref{staro1} (a) are displayed some solutions of Eq. (\ref{phireq}) for $R$ as a function of the exponent $n$ and for some values of $\phi$.

\begin{figure}
\centering
\subfigure[]{
\resizebox*{4cm}{!}{\includegraphics{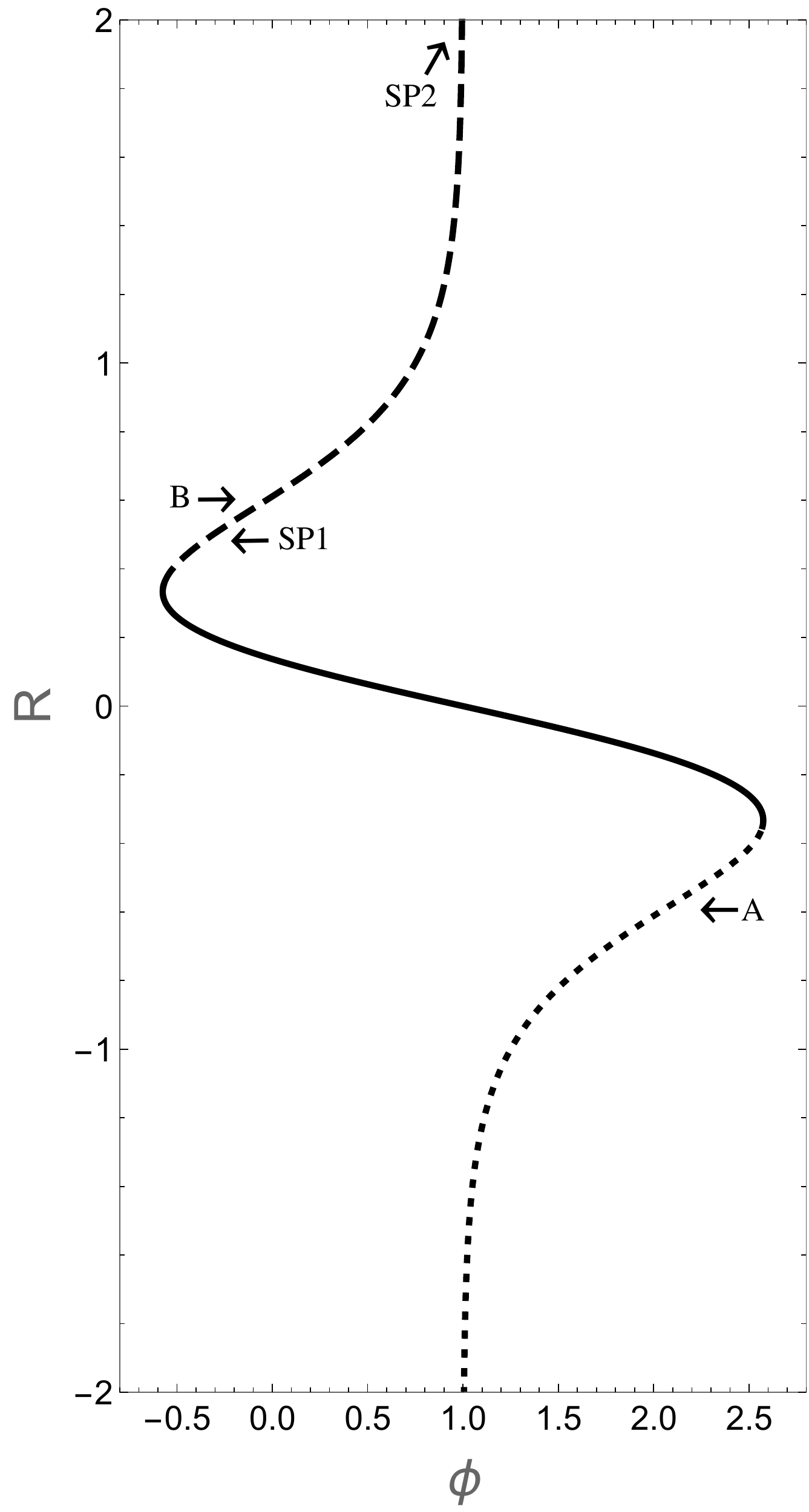}}}\hspace{5pt}
\subfigure[]{
\resizebox*{4.15cm}{!}{\includegraphics{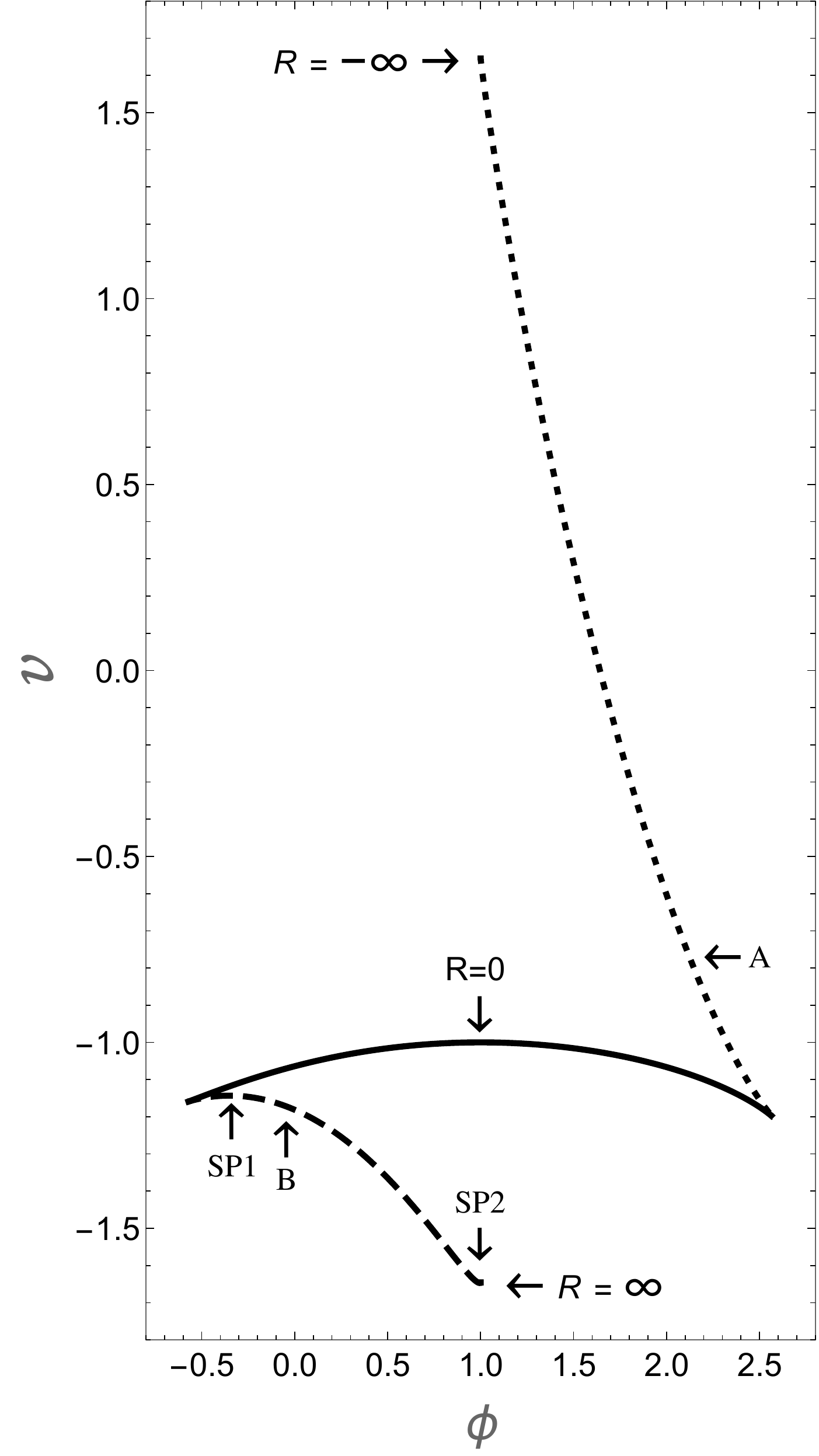}}}\hspace{10pt}
\subfigure[]{
\resizebox*{4cm}{!}{\includegraphics{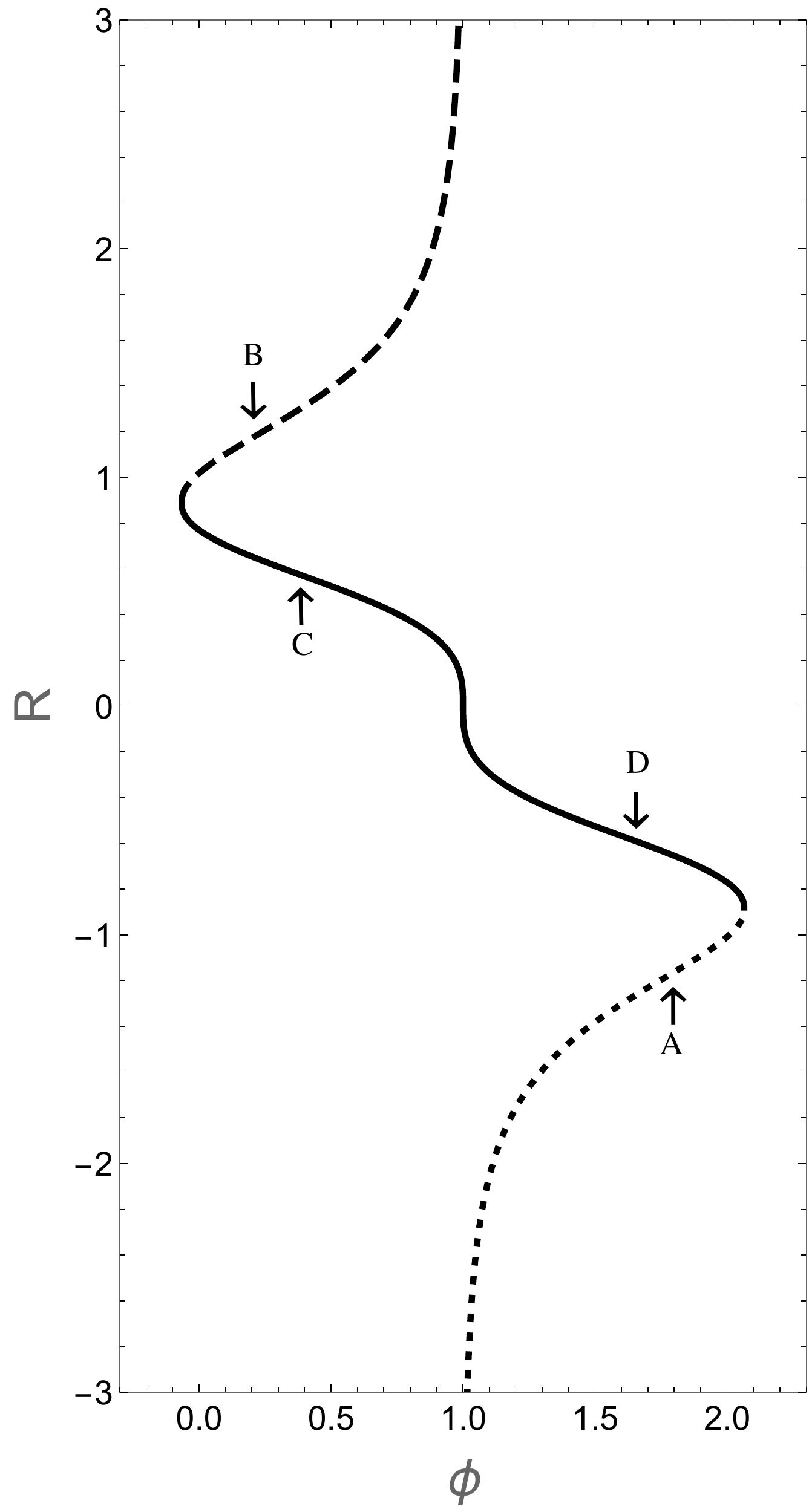}}}\hspace{5pt}
\subfigure[]{
\resizebox*{4.15cm}{!}{\includegraphics{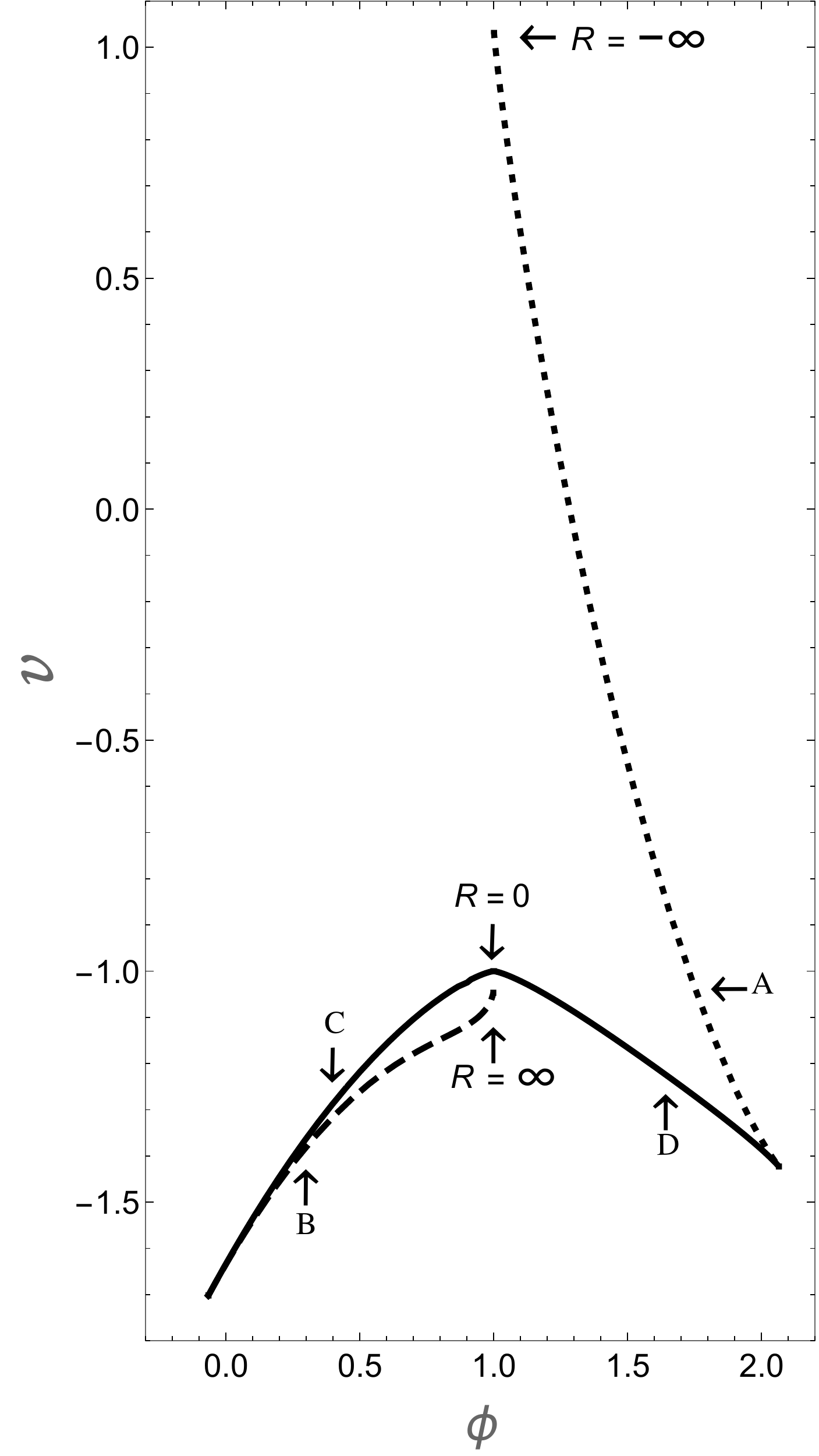}}}
\\
\subfigure[]{
\resizebox*{4cm}{!}{\includegraphics{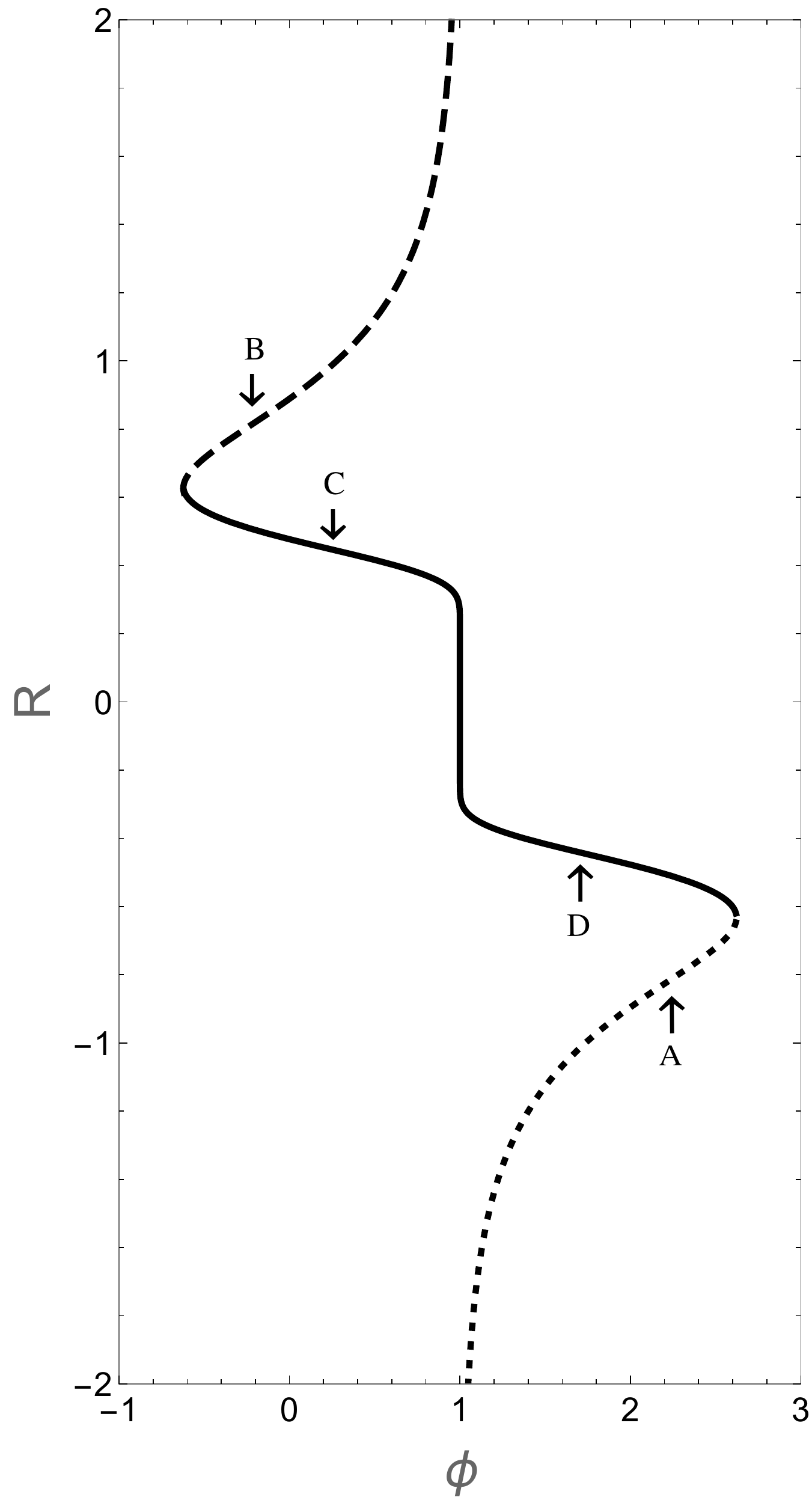}}}\hspace{5pt}
\subfigure[]{
\resizebox*{4.15cm}{!}{\includegraphics{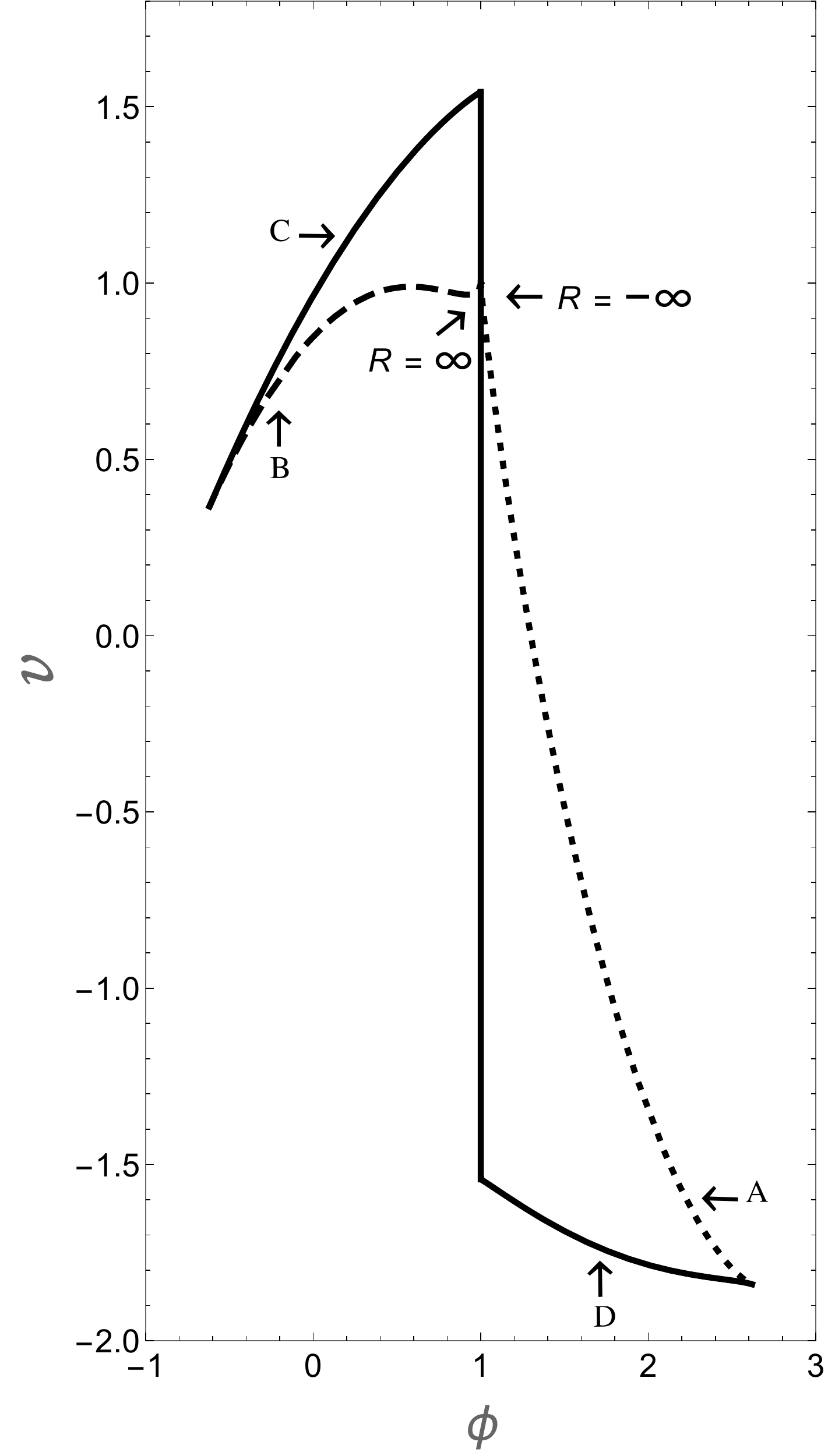}}}\hspace{10pt}
\subfigure[]{
\resizebox*{4cm}{!}{\includegraphics{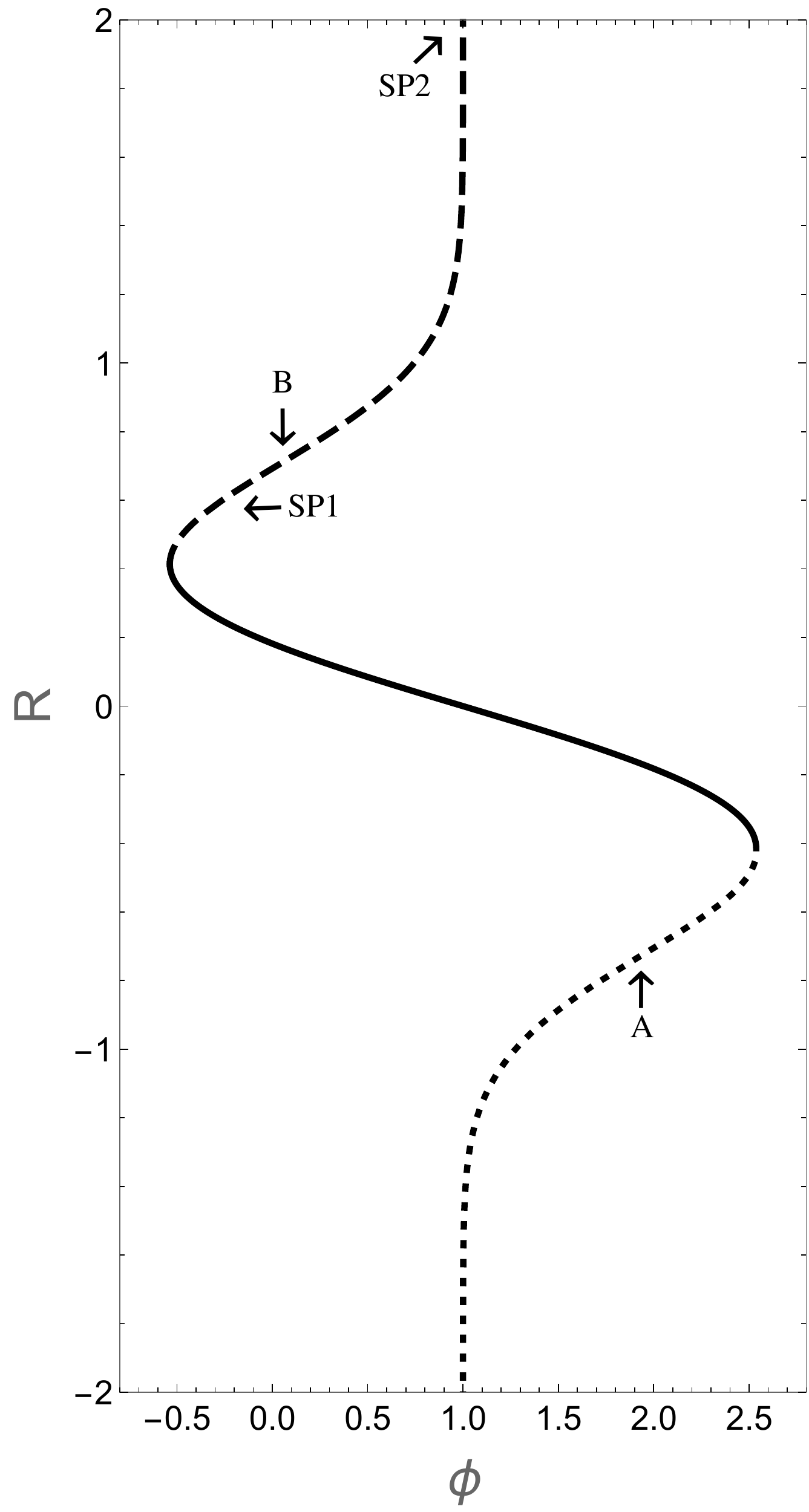}}}\hspace{5pt}
\subfigure[]{
\resizebox*{4.15cm}{!}{\includegraphics{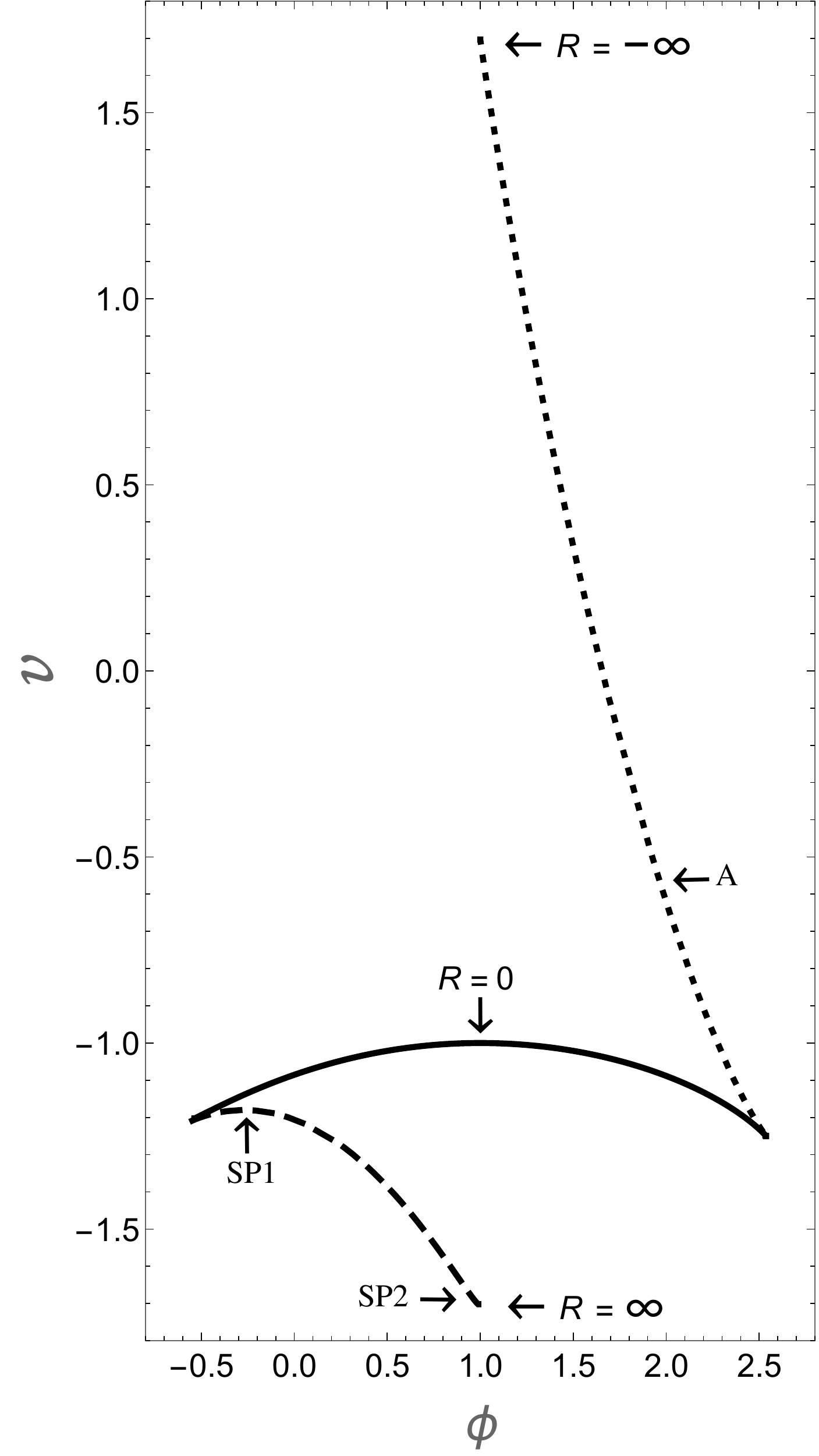}}}
\caption{Scalar curvature (left) and effective potential (right) as a function of the scalar field for the Starobinsky, (a) and (b), Hu-Sawicki, (c) and (d), complementary exponential, (e) and (f), and Modified Starobinsky, (g) and (h), models; with $\mu=R_0=1$ and $n=4$ ($n=1$ in the complementary exponential), lines are matched for each pair of figures. For each model it is shown the mapping of points at $R\to\pm\infty$ to the points showed in the right panel. Starobinsky model exhibit other seven points of interest: absolute maximum and minimum $\phi=-0.574$ and $\phi=2.574$ which are reached in $R=\pm 1/3$ respectively, $\phi$=-0.096 (A) and 2.096 (B) at $R=\pm0.057$, the de Sitter (in vacuum $T=0$) non stable points SP1, $R=0.484$ and $R=0$, and the stationary point SP2, $R=1.985$; $R=0$ is an inflection point too in (a), de Sitter points can also be found as a local maxima of the effective potential. The scalar field in the Hu-Sawicki model (c) presents two maximum $\phi$=-0.065 and 2.065 at $R=\pm0.880112$, respectively; in this model, as well as the complementary exponential, two other inflection points appear, C and D, however, due to the exponent in this case $n=4$, this model only presents one non stable de Sitter point $R=0$ and five inflection points in the marks A, B, C, D, and in $R=0$, which are  $\phi$=-2.377 and $-2.37764$ at $R=\pm0.570017$ , however, its effective potential have and  the Hu-Sawicki model. Modified Starobinsky effective potential (h) has a geometry similar to (b), and thus the same number of characteristic points. Modified Starobinsky scalar field presents some region with $\phi=0$ between maximal points like the Hu-Sawicki model for certains exponent values, Fig. \ref{fig:staro3}, so that the effective potential will not have a continuos derivative at $R=0$.}\label{fig:severalplots}
\end{figure}

The effective potential (\ref{potentialv}) for this model is calculated as
\begin{multline}
    v(R,n,\mu)=\frac{\mu ^2 n R}{\left(R^2+1\right)^{n+1}} \left(4-\frac{(2 n+1) R^2+1}{\mu  n R}-\frac{[n (4 n+7)+4] R^2+5 n+4}{(2 n+1) \left(R^2+1\right)^{n+1}}\right)\\-\frac{3 \mu ^2 n^2R}{2 n+1} \, _2F_1\left(\frac{1}{2},2 n+1;\frac{3}{2};-R^2\right),
\end{multline}
which is plotted in Fig \ref{fig:severalplots} (b) from those points that solve Eq. (\ref{phireq}), shown in the same figure (a). In this plot it is possible to appreciate the mapping of points at infinity in plane $\phi-R$ to points in the plane $\phi-v$ with finite distances [\cite{frolov2}]. The position of the two maximal and the three inflection points in $R$ do not depend on $\mu$ and are in $\pm R_0/\sqrt{2 n+1}$, and $R=\pm R_0\sqrt{3 (n+1)/(2 n^2+3 n+1)}$ and $R=0$ ($\phi=1$), respectively, whose values $\phi$ do not depend on $R_0$
\begin{equation}
\phi=1\mp\frac{2\mu n}{\sqrt{2 n+1}}\left(\frac{1}{2 n+1}+1\right)^{-n-1}\qquad\qquad\text{and}\qquad\qquad \phi=1\mp \frac{\mu  n \sqrt{6 n+3}}{n+2} \left(\frac{3}{2 n+1}+1\right)^{-n}.
\end{equation}

\begin{figure}[ht]
\centering
\subfigure[]{
\resizebox*{8.5cm}{!}{\includegraphics{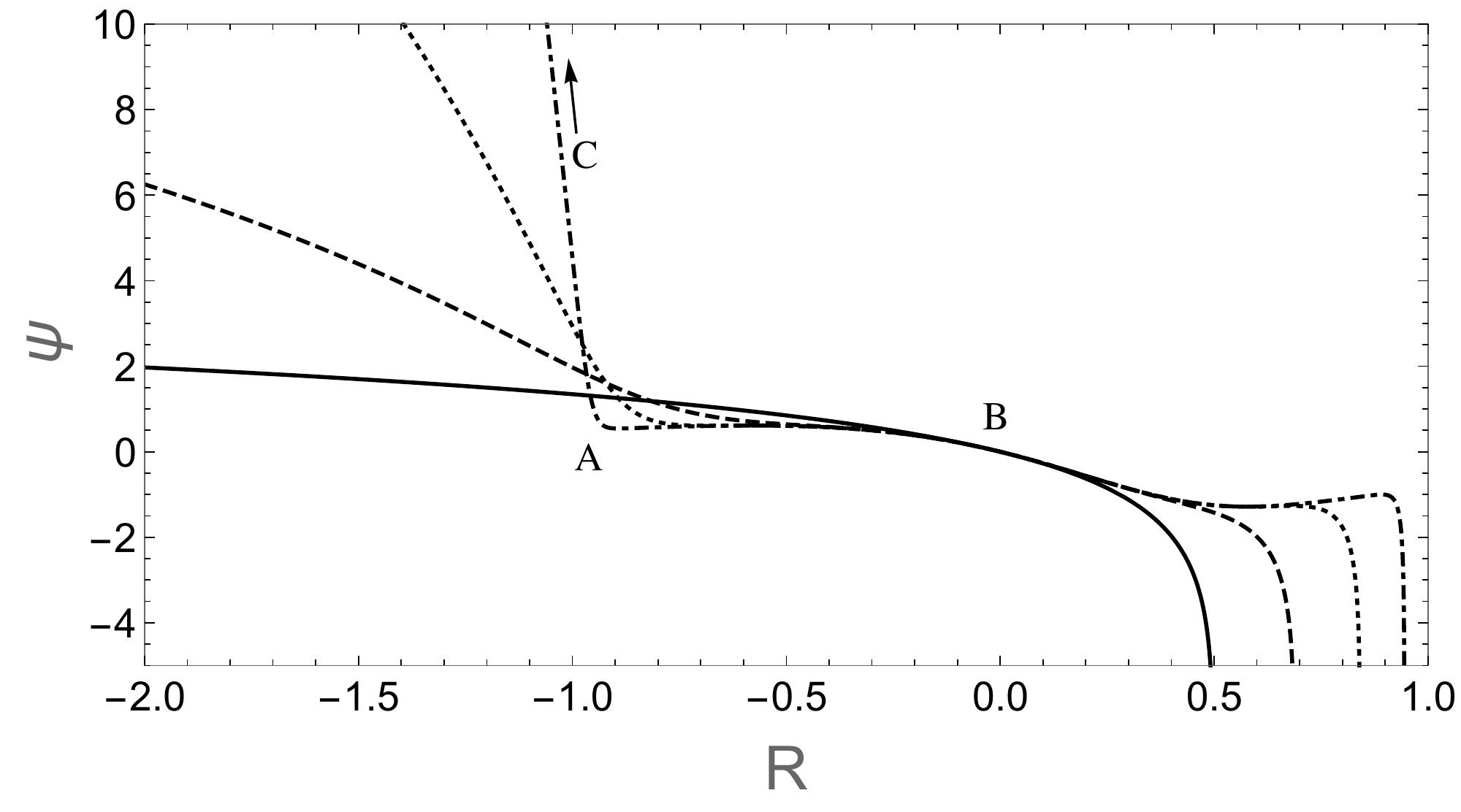}}}\hspace{5pt}
\subfigure[]{
\resizebox*{8.5cm}{!}{\includegraphics{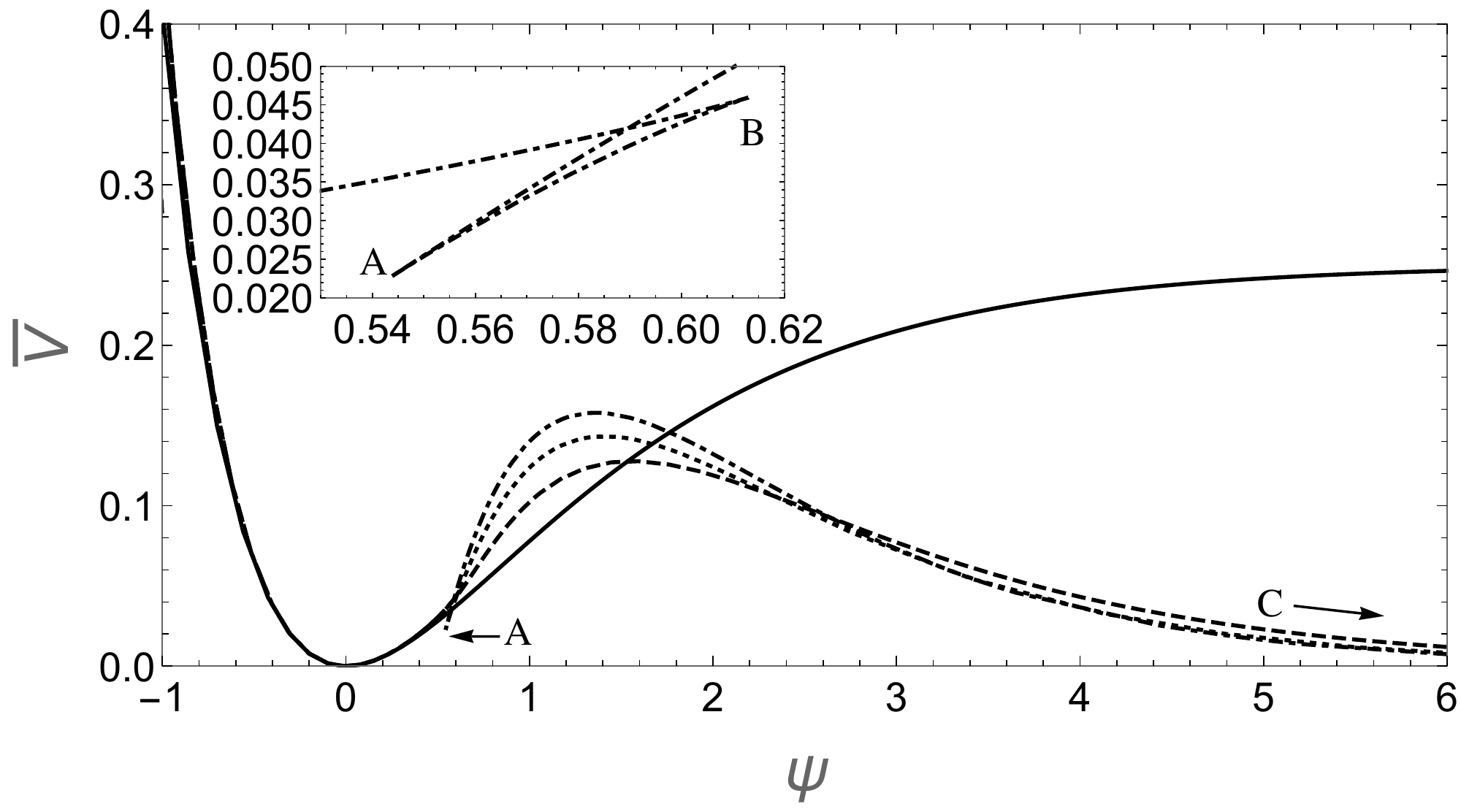}}}
\\
\subfigure[]{
\resizebox*{8.5cm}{!}{\includegraphics{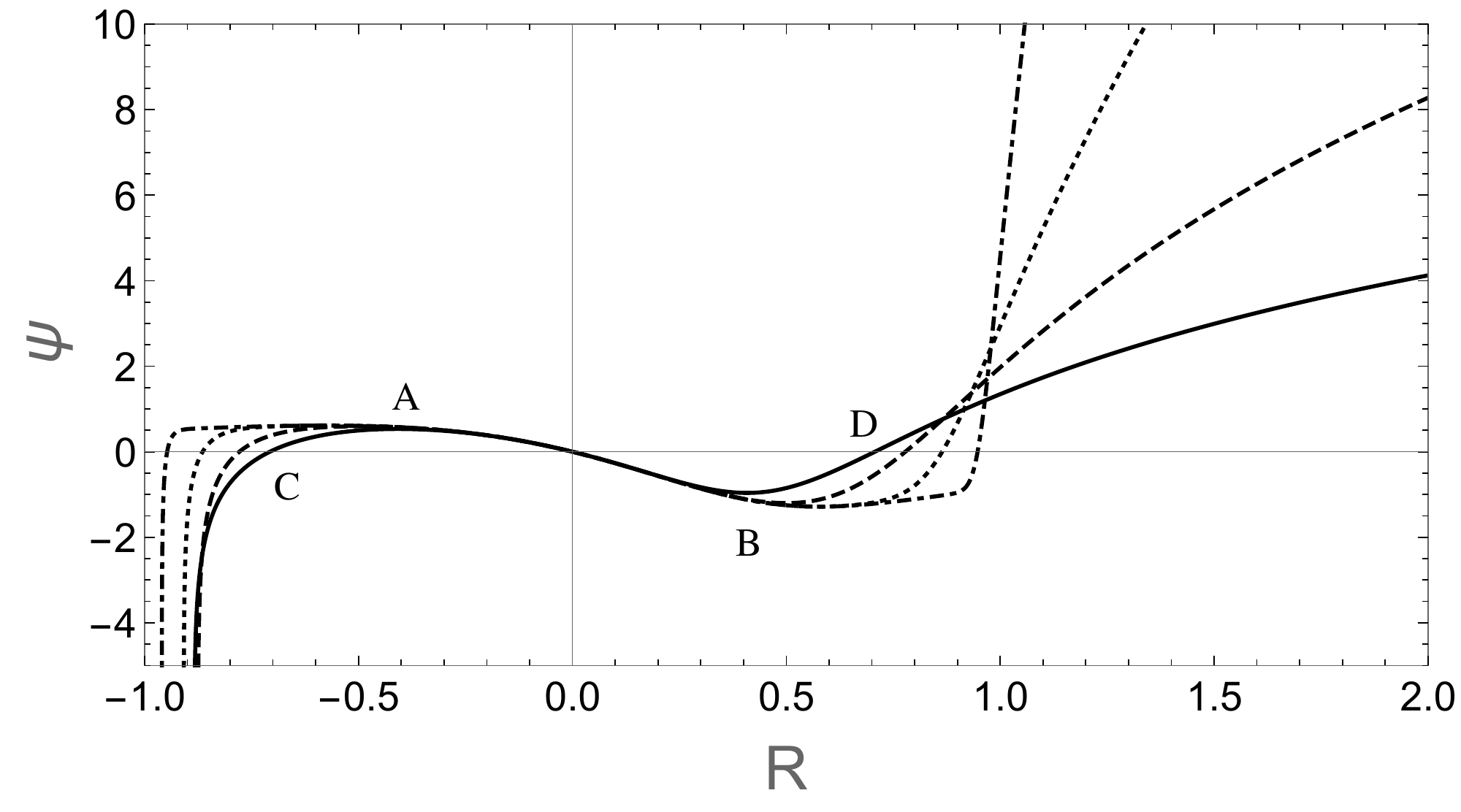}}}\hspace{5pt}
\subfigure[]{
\resizebox*{8.5cm}{!}{\includegraphics{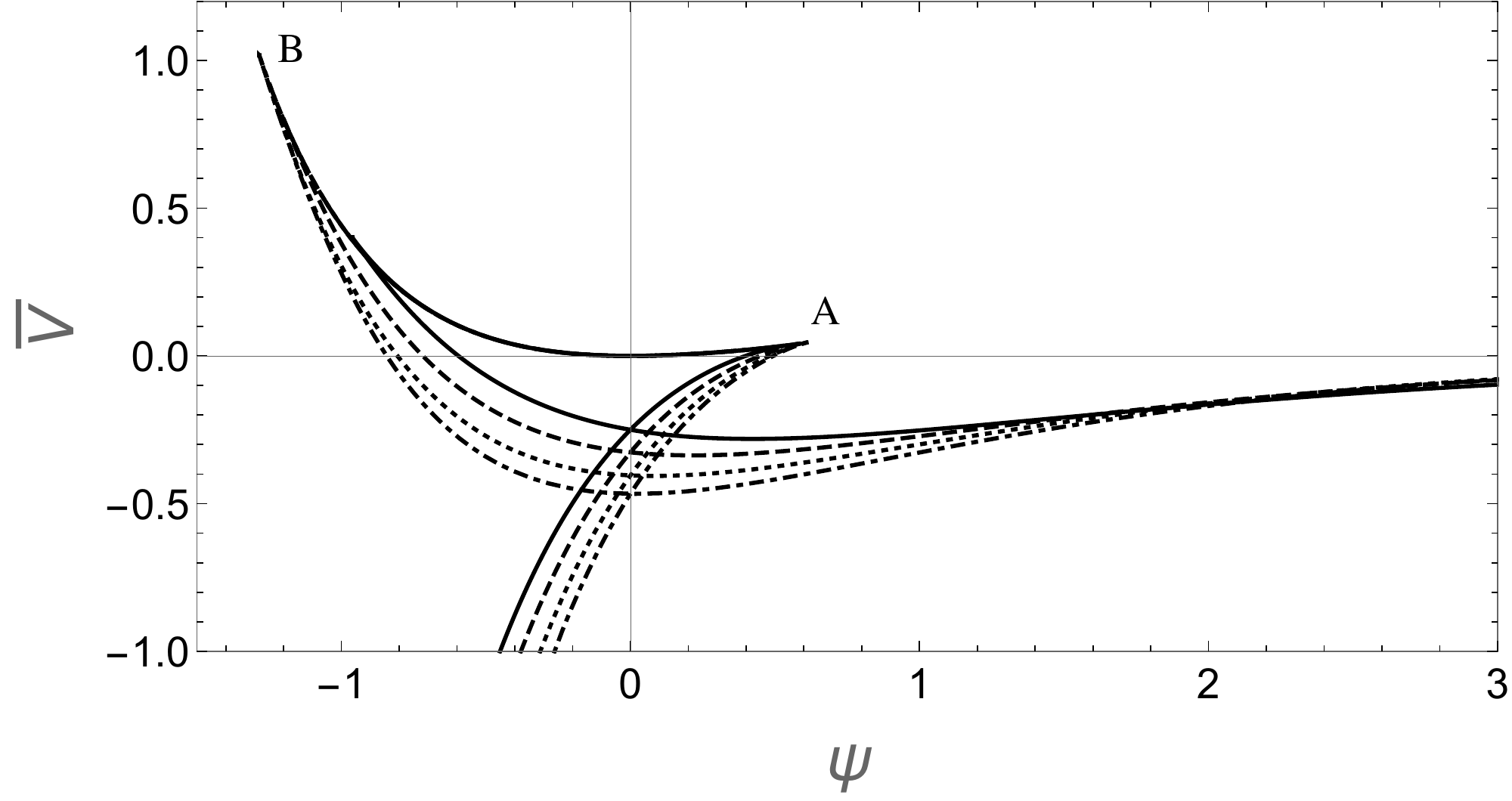}}}
\caption{Scalar field $\psi$ as a function of $R$ (left) and scalar field potential $\tilde{V}$ as a function of $\psi$ (right) for $p=1, 3, 9, 39$ (a) and (b), and for $p=2, 4, 10, 40$ (c) and (d), Black, dashed, dotted, dot-dashed lines; for the Starobinsky model expansion at $R=0$ to order $p$. When $p=1$ inflationary Starobinsky potential [\cite{Starobinsky:1980te}] is obtained. Uppercase letters show mapping between regions. In (a) it is observed that at A there is an inflection point which is translated into (b) that from points at infinity to $A$, the dominant term is $R^2$. In (c) and (d) there are four points of interest, A and B, maximal points, which imply a change of direction in the potential, and C and D, roots, which translate into the intersection points at $\psi=0$ in (d). In all panels $\mu=n=1$.} \label{fig:staro1}
\end{figure}

\begin{figure}[ht]
\centering
\subfigure[]{
\resizebox*{8.5cm}{!}{\includegraphics{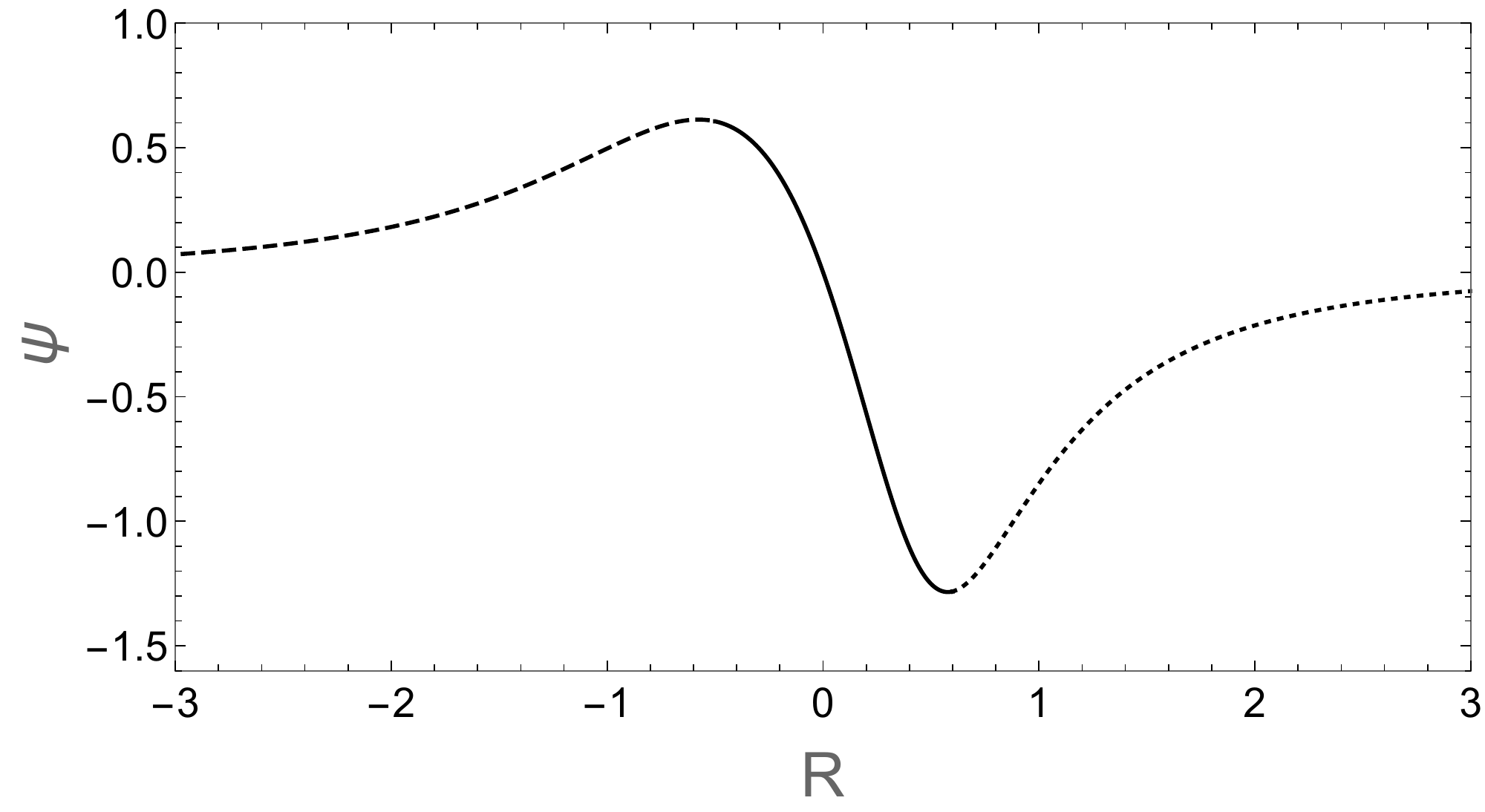}}}\hspace{5pt}
\subfigure[]{
\resizebox*{8.5cm}{!}{\includegraphics{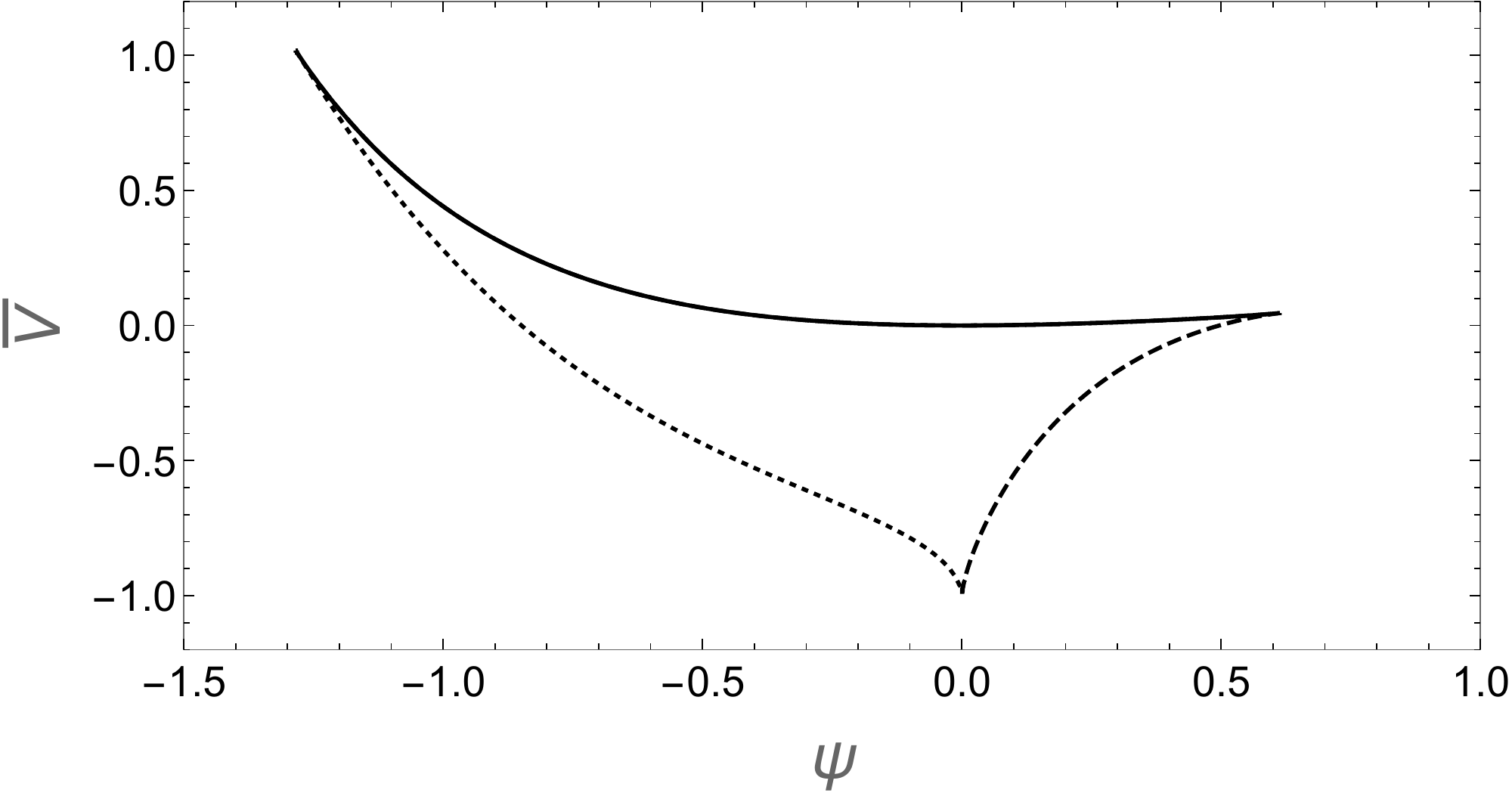}}}
\caption{Scalar field $\psi$ as a function of $R$ (a) and scalar field potential $\tilde{V}$ as a function of $\psi$ (b) for the Starobinsky model (\ref{staro2007}).} \label{fig:staro2}
\end{figure}

Starobinsky model has a fairly important physical content, since its expansion at $R=0$ to order $p$ reproduces the $R^2$ term of its own famous 1980 model [\cite{Starobinsky:1980te}]. From
\begin{equation}\label{staroexpansion}
    f(R)=R-\frac{\mu  n R^2}{R_0}+\mu  n \sum _{m=2}^{p} \frac{(-1)^m(n+1)_{m-1}}{m!} \frac{R^{2 m}}{R_0^{2 m-1}},
\end{equation}
where $(n+1)_{m-1}$ is the Pochhammer symbol, it is possible to see the form of potential (\ref{potentialvtilde}) in the Einstein frame, through transformation (\ref{transformation1}), but first we need to find the scalar field $\psi$ as a function of $R$, that is
\begin{equation}\label{scalarpotentialstaro}
    \psi(R)=\frac{3}{2}\ln\left[1-2 \mu  n \frac{R}{R_0}+2 \mu  n \sum _{m=2}^p (-1)^m \frac{(n+1)_{m-1}}{\Gamma (m)} \left(\frac{R}{R_0}\right)^{2 m-1}\right],
\end{equation}
note that only for $p=2$, the scalar curvature can be obtained analytically in terms of $\psi$, likewise, the Eq. (\ref{scalarpotentialstaro}) as well as its derived scalar potential (\ref{potentialvtilde}) depend strongly on the parity of the order of the expansion, Fig. \ref{fig:staro1} (a) and (b) for odd $p$, an (c) and (d) for even $p$. When $p\to\infty$ the total form of model (\ref{staro2007}) is recovered, and the scalar field and potential are plotted in Fig. (\ref{fig:staro2}), where the similarity with the potential of Fig. \ref{fig:staro1} (d) can be observed.

Finally, in Fig. \ref{scalarpotentials} it is shown the scalar potential (\ref{scalarpotential1}) as a function of $\phi$ for $\mu=1$.

\subsection{Hu-Sawicki model}
Hu and Sawicki [\cite{Hu:2007nk}] proposed a power law model to describe accelerated expansion without a cosmological constant and satisfies both, cosmological and Solar systems tests in the small-field limit. This model is given by
\begin{equation}\label{husawicki}
    f(R)=R-\mu \frac{c_1(R/\mu)^{p}}{1+c_2(R/\mu)^{p}},
\end{equation}
where $\mu,c_1,c_2,p>0$ are the parameters of the model. If $c_1=c_2=1$ are set, equation $F-\phi=0$ takes the form 
\begin{equation}\label{husawieq}
    (1-\phi)\left[\left(\frac{R}{\mu }\right)^p+1\right]^2-p \left(\frac{R}{\mu}\right)^{p-1}=0,
\end{equation}
not for all values of $R$ and $p$ this equation can be solved, since it is observed that there will be divergences in $R=\mu (-1)^{1/p}$; however, some solutions are plotted in Fig. \ref{fig:staro3} (b), where it is noted the singularity at $R=-1$ when $n=1$.
The relation of $R$ with respect to $\mu$ and $\phi$, for $p=1$ is given by
\begin{equation}
    -\mu ^2 \phi +R^2 (1-\phi )+2 \mu  R (1-\phi )=0,
\end{equation}
and the solution for $\phi$ in terms of $\mu$ is a straight line with slope $\pm \left(1-(1-\phi)^{-1/2}\right)$ for $1>\phi$. Similarly, Eq. (\ref{husawieq}) can be solved for $p$ with $\mu=1$, and for some values of $\phi$, as depicted in Fig. \ref{staro1} (b).

In Fig. \ref{fig:severalplots} the points satisfying $F-\phi=0$ (c) are observed for function (\ref{husawicki}), and their image through the effective potential
\begin{multline}
    v(R,p,\mu)=\frac{p}{4 R \left(R^p+1\right)} \left(\frac{(1-4 R) R^p-4 R}{p \left(R^p+1\right)}+\frac{\left(4 (2-R) R^p-4 R+3\right) R^p+1}{\left(R^p+1\right)^2}-\frac{2 p R^{2 p}}{\left(R^p+1\right)^3}-\frac{1}{p^2}\right)\\+\frac{p}{4 R} \left(\frac{1}{p^2}-1\right) \, _2F_1\left(1,-\frac{1}{p};\frac{p-1}{p};-R^p\right),
\end{multline}
are shown in the same Figure (d), and in Fig. \ref{scalarpotentials} (b) is the numerical plot of the scalar potential.

For Hu-Sawicki and Starobinsky models in general, scalar potential can not be expressed analytically as a function of all parameters and the scalar curvature, however this is not be true for the next two models as will be seen below.

\subsection{Tsujikawa model}
In order to satisfy local gravity constraints as well as conditions of the cosmological scenario, Tsujikawa proposed the model [\cite{Tsujikawa:2007xu}]
\begin{equation}\label{tsujikawa}
    f(R)=R-\mu R_T \tanh{(R/R_T)},
\end{equation}
with $\mu, R_T>0$ the parameters of the model. For this model it is possible to invert $F-\phi=0$ analytically and find the potentials $v$ and $V$. For this we see that
\begin{equation}
    R=R_T\arccosh\pm \sqrt{\frac{\mu }{1-\phi }},
\end{equation}
such that $1-\mu\leq \phi< 1$ to $R\in\Re$. The effective potential, Eq. (\ref{potentialv}), associated is
\begin{multline}
    v(R)=\frac{\mu}{8\cosh^4{\left(R/R_T\right)}} \left[(1-\mu) R_T \sinh \left(\frac{4 R}{R_T}\right)+2 (1+2 \mu) R_T \sinh \left(\frac{2 R}{R_T}\right)\right.\\ \left.-4 R \cosh \left(\frac{2 R}{R_T}\right)-4 (1+\mu) R\right],
\end{multline}
and in terms of the potential $\phi$
\begin{equation}
    v(\Phi)=\frac{\mu R_T}{4 \Phi^4} \left\{\left[3\mu+2(1-\mu) \Phi^2 \right]\sinh \left[2 \arccosh(\pm \Phi)\right]-2 (\mu +2 \Phi^2 ) \arccosh{(\pm\Phi)}\right\},
\end{equation}
where $\Phi=\sqrt{\mu/(1-\phi)}$, with which is possible to write the scalar potential through the integral (\ref{scalarpotential1}), giving
\begin{equation}
    V(\Phi)=\frac{\mu R_T}{\Phi^2}\left[\Phi\sqrt{\Phi^2-1}-\arccosh(\pm\Phi)\right].
\end{equation}
Figure \ref{potentials} (a) and (c) shows the effective and scalar potentials for some values of $\mu$.

\subsection{Exponential model}
\begin{figure}[ht]
\centering
\subfigure[]{
\resizebox*{8.5cm}{!}{\includegraphics{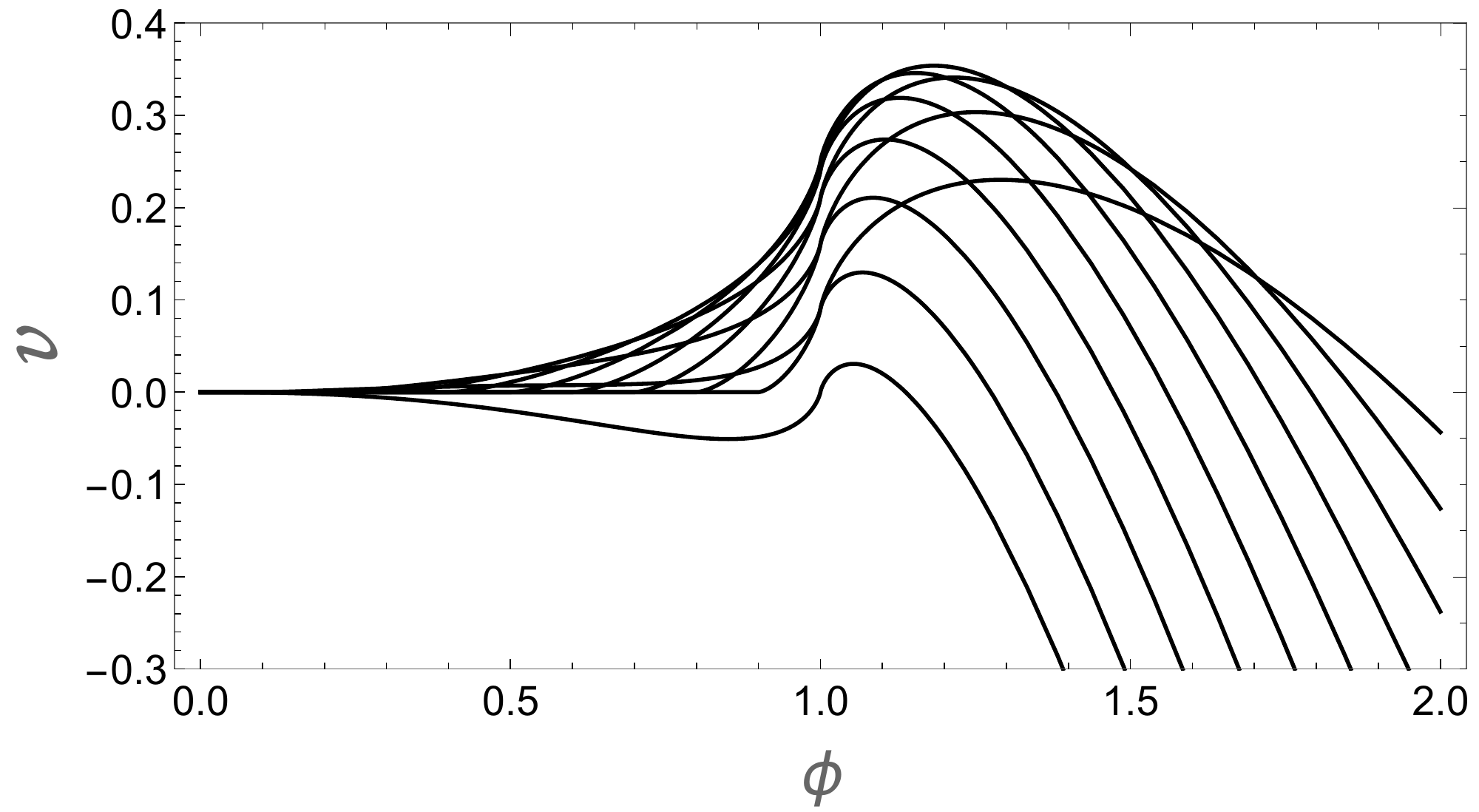}}}\hspace{5pt}
\subfigure[]{
\resizebox*{8.5cm}{!}{\includegraphics{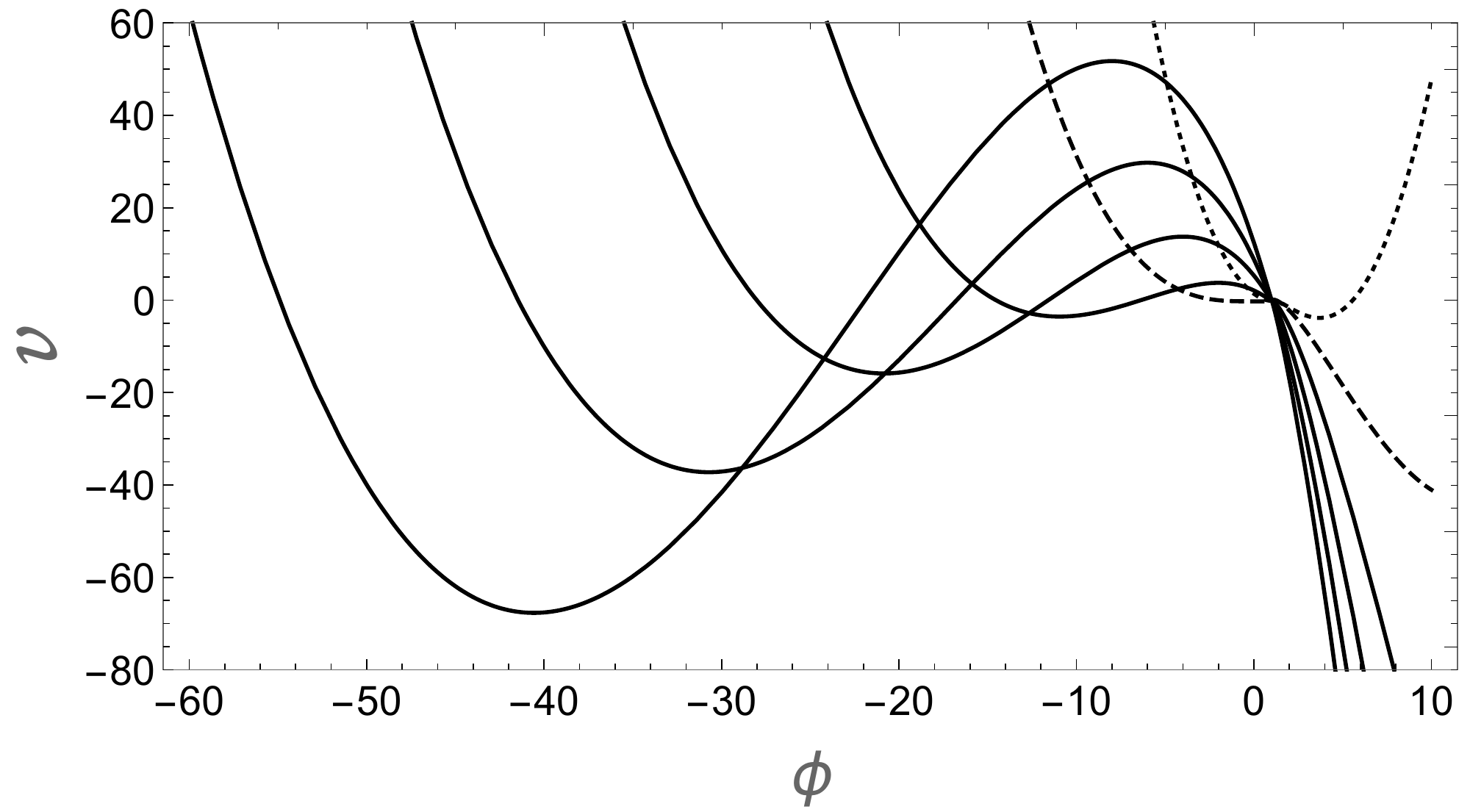}}}
\\
\subfigure[]{
\resizebox*{8.5cm}{!}{\includegraphics{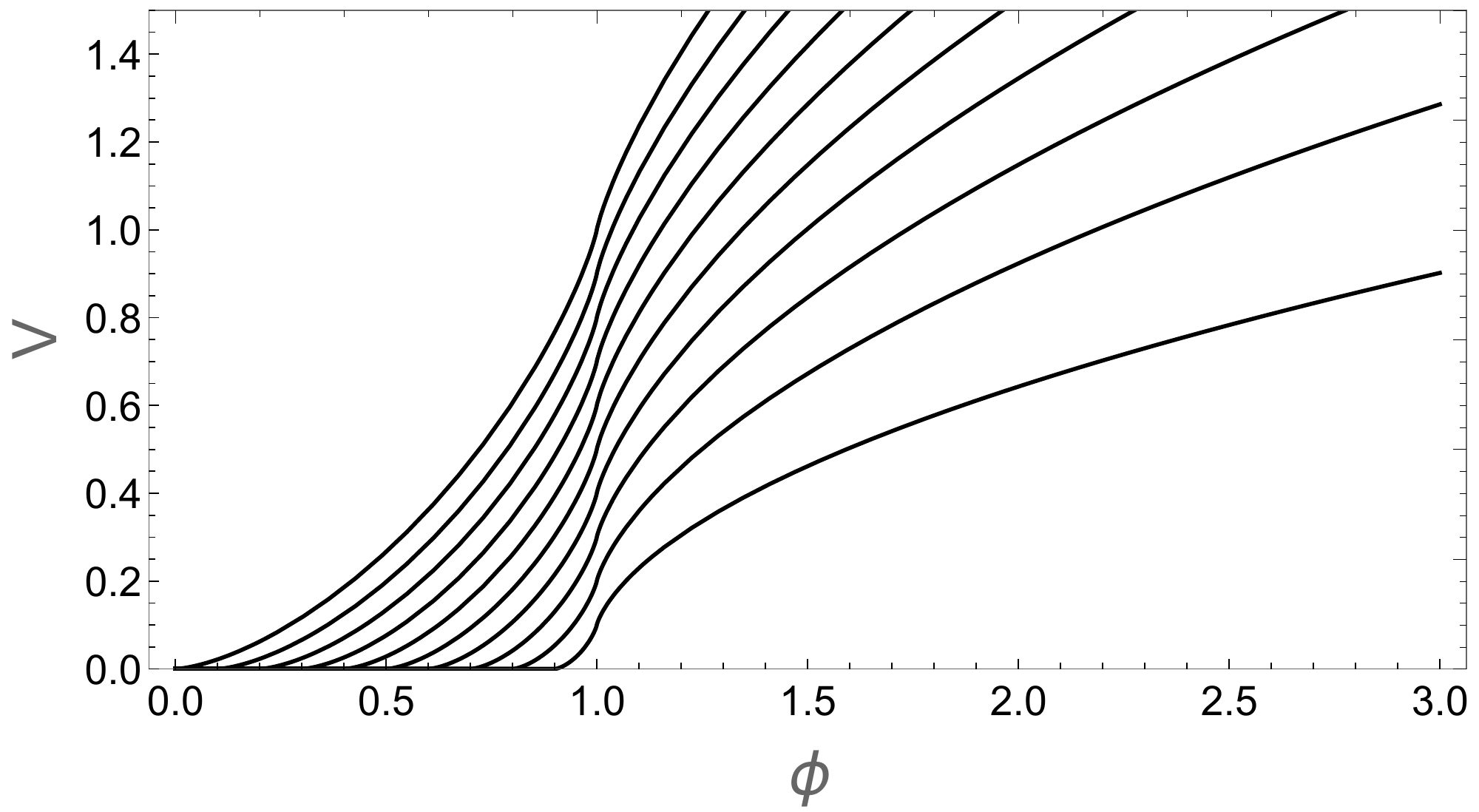}}}\hspace{5pt}
\subfigure[]{
\resizebox*{8.5cm}{!}{\includegraphics{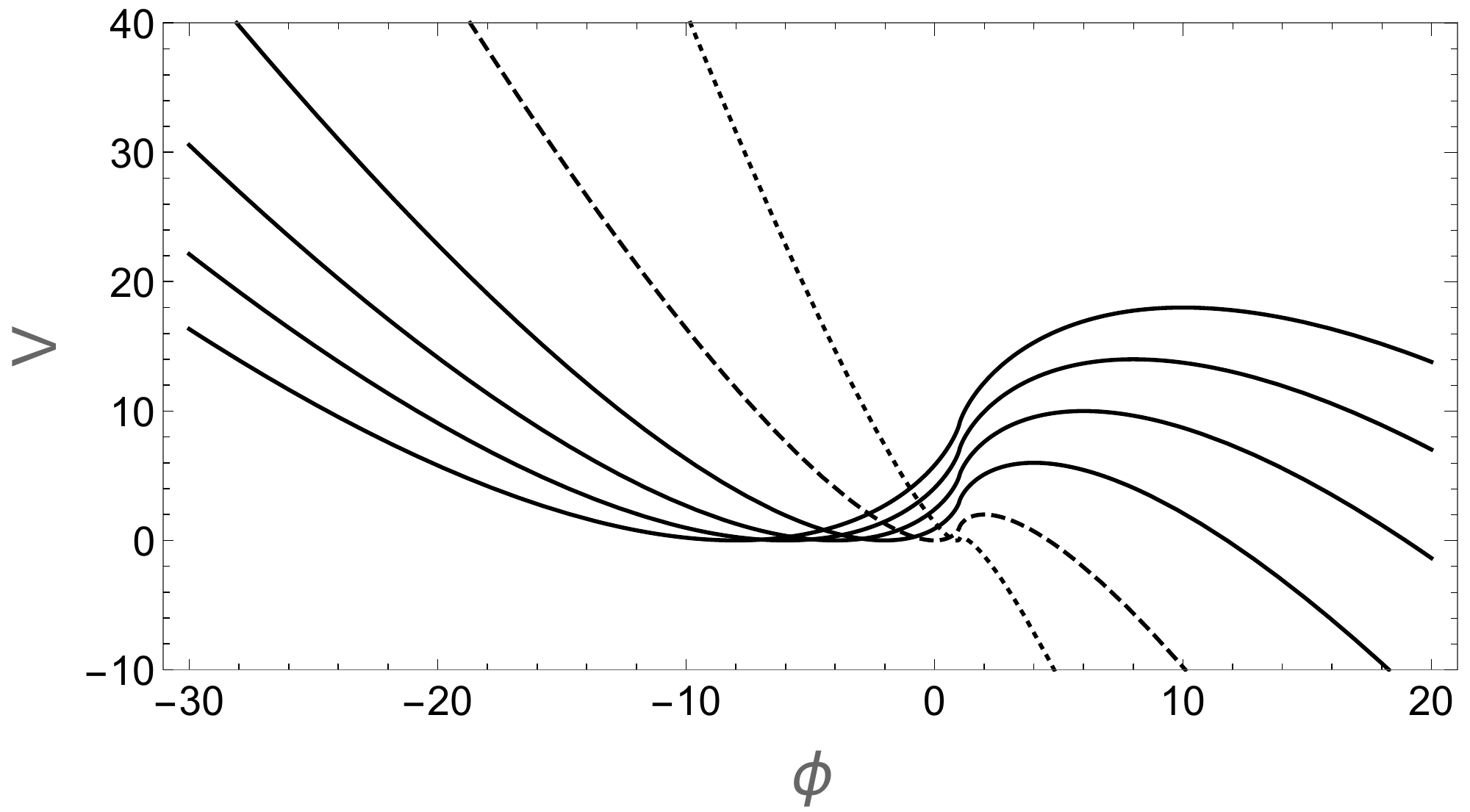}}}
\caption{Potential $v(\phi)$ (a) and (b) and scalar potential $V(\phi)$, (c) and (d), for the Tsujikawa (\ref{tsujikawa}), (a) and (c) and exponential (\ref{exponential}) models, (b) and (d). In (a) from left to right and bottom to top, and in (c) from left to right, $\mu=0.1, 1, 3, 5, 7, 9$. In (b) and (d) from left to right $\mu=9, 7, 5, 3, 1, 0.1$. Although all potential are continuous functions, in the interval $1\leq\phi$, both models have complex values and the real part has been plotted.}\label{potentials}
\end{figure}

Another interesting model that explains accelerated expansion constructed to mimic $\Lambda$CDM Universe and able to pass the Solar systems tests, was proposed in the middle of the last decade for different authors [\cite{Zhang:2005vt,Bean:2006up,Cognola:2007zu}]. In this model, $f(R)$ takes the form
\begin{equation}\label{exponential}
    f(R)=R-\mu R_E\left(1-e^{-R/R_E}\right),
\end{equation}
where $\mu$ and $R_E$ are the parameters of the model. If $\mu>0$ and $\phi<1$ or $\mu<0$ and $\phi>1$, the Ricci scalar can be written in terms of $\phi$, that is
\begin{equation}
    R=R_E\ln{\left(\frac{\mu}{\phi-1}\right)},
\end{equation}
and the effective potential is given by
\begin{equation}
    v(R)=-\frac{1}{4} \mu  e^{-\frac{2 R}{R_E}} \left(\mu  (2 R+5 R_E)+4 e^{R/R_E} (R-2 \mu  R_E+R_E)\right),
\end{equation}
or in terms of $\phi$
\begin{equation}
    v(\phi)=\frac{1}{4} R_E (1-\phi) \left(2 (\phi -3) \ln \left(\frac{\mu }{1-\phi}\right)+8 \mu +5 \phi -9\right),
\end{equation}
and the scalar potential
\begin{equation}
    V(\phi)=R_E \left[(\phi -1) \ln \left(\frac{\mu }{1-\phi}\right)+\mu +\phi -1\right].
\end{equation}
The functional form of these potentials can be seen in Fig. \ref{potentials} as a function of $\phi$ and $\mu$.

\subsection{Complementary exponential model}
Let us suppose the following model,
\begin{equation}\label{anotherexp}
    f(R)=R-\mu \left(1+\frac{R_0^2}{R^2}\right)e^{-n R_0^2/R^2},    
\end{equation}
where $R_0$ is the constant scalar curvature in vacuum $R_0=2\mu(3-2n)e^{-n}$, and $\mu, n$ parameters of the model, conditioned by (\ref{condition2}) as ${\mu}\in \mathbb{R}$ and $n>0$; and  (\ref{condition1}) implying
\begin{equation}\label{limit}
    \lim_{R \to \infty}\tilde{f}(R)=-\mu.
\end{equation}
As shown in [\cite{Perez-Romero:2017njc}], a model expressed in the form
\begin{equation}\label{flambda}
    \tilde{f}(R)=-2\Lambda y(R,b),
\end{equation}
turns out to be an extension of the the $\Lambda$CDM model if the cosmological constant is reproduced at some limit of $b$, which is a characteristic parameter of each model. For (\ref{anotherexp}) it is found
\begin{equation}
    y(R,b)=\left[1+\left(\frac{b \Lambda}{R}\right)^2\right]e^{-n (b \Lambda/R)^2},
\end{equation}
with $\Lambda=\mu/2$ and $b=2R_0/\mu$, and the equivalency with the $\lambda$CDM model seen through the limit (\ref{limit}) written as
\begin{equation}
    \lim_{b \to 0}f(R)=R-2\Lambda,
\end{equation}
and 
\begin{equation}
    \lim_{b \to \infty}f(R)=R.
\end{equation}
Model (\ref{anotherexp}), as well as Starobinsky and Hu-Sawicki models can be expressed in the form (\ref{flambda}) and therefore recover the viability of the $\Lambda$CDM model. For this model the relation between $R$ and $\phi$ is
\begin{equation}\label{rphimym}
    R^5 (1-\phi ) e^{n/R^2}-2 \left((n-1) R^2+n\right)=0,
\end{equation}
which, in general has no analytical solutions, however Fig. \ref{fig:staro3} (c) shows $\phi$ as a function of $R$ and the power $n$ for $\mu=R_0=1$. By calculating the integral (\ref{potentialv}) it is found
\begin{multline}
    v(x)=e^{-2 n/x^2} \left\{\frac{1}{R_0 x}\left[-\frac{9 (8 n (2 n+1)+11)}{128 n^2}+\frac{8 n (10 n-19)-33}{32 n x^2}-\left(n-\frac{15}{4}\right) \left(2 n-\frac{3}{2}\right)\frac{1}{x^4}\right.\right.\\
    \left.\left.+\left(\frac{13}{2}-4 n\right)\frac{ n}{x^6}-\frac{2 n^2}{x^8}\right]+\left(1+\frac{3-2 n}{x^2}-\frac{2 n}{x^4}\right)e^{n/x^2}\right\}+9\sqrt{2 \pi }\frac{8 n (2 n+1)+11}{512 R_0n^{5/2}}\erf\left(\frac{\sqrt{2 n}}{x}\right)
\end{multline}
where $x=R/R_0$. Fig \ref{fig:severalplots} (f) shows the relation between $v$ and $\phi$, from the solution of Eq. (\ref{rphimym}), plotted in the panel (e) of the same Figure, and in Fig. \ref{scalarpotentials} (c) the shape of the scalar potential is observed in terms of the scalar field $\phi$ and $n$.
\subsection{Modified Starobinsky model}
\begin{figure}
\centering
\subfigure[]{
\resizebox*{7cm}{!}{\includegraphics{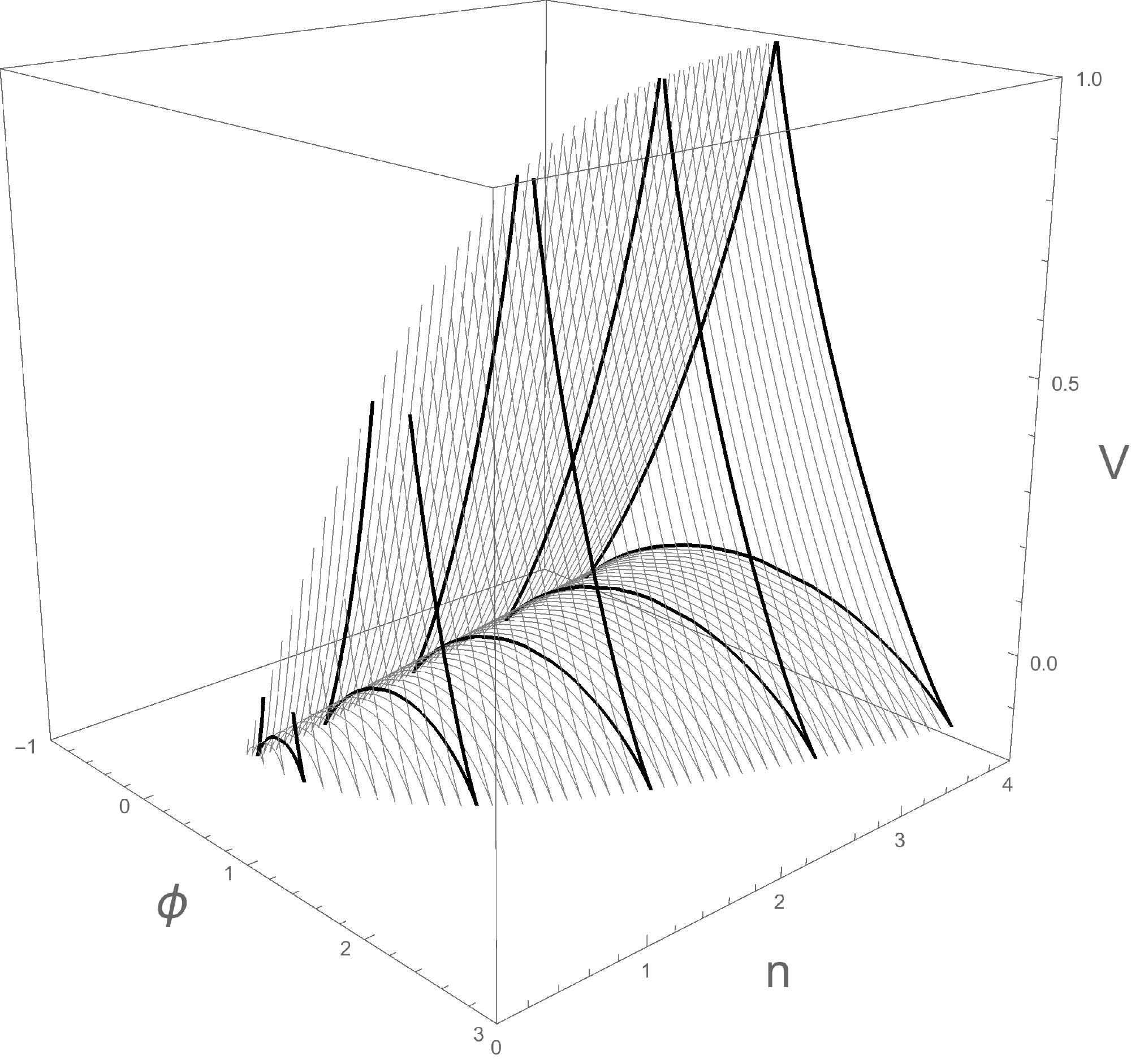}}}\hspace{5pt}
\subfigure[]{
\resizebox*{7cm}{!}{\includegraphics{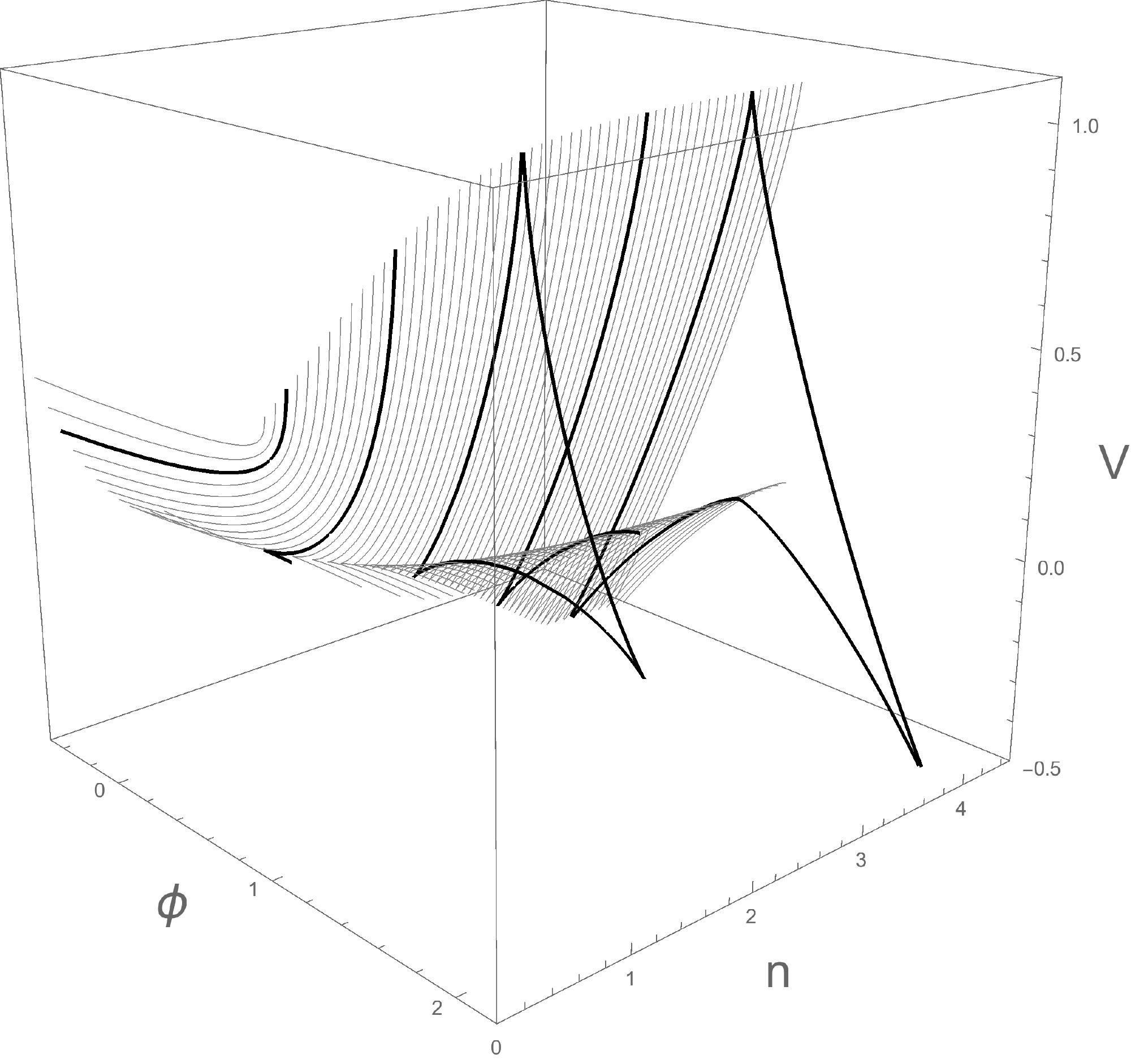}}}
\\
\subfigure[]{
\resizebox*{7cm}{!}{\includegraphics{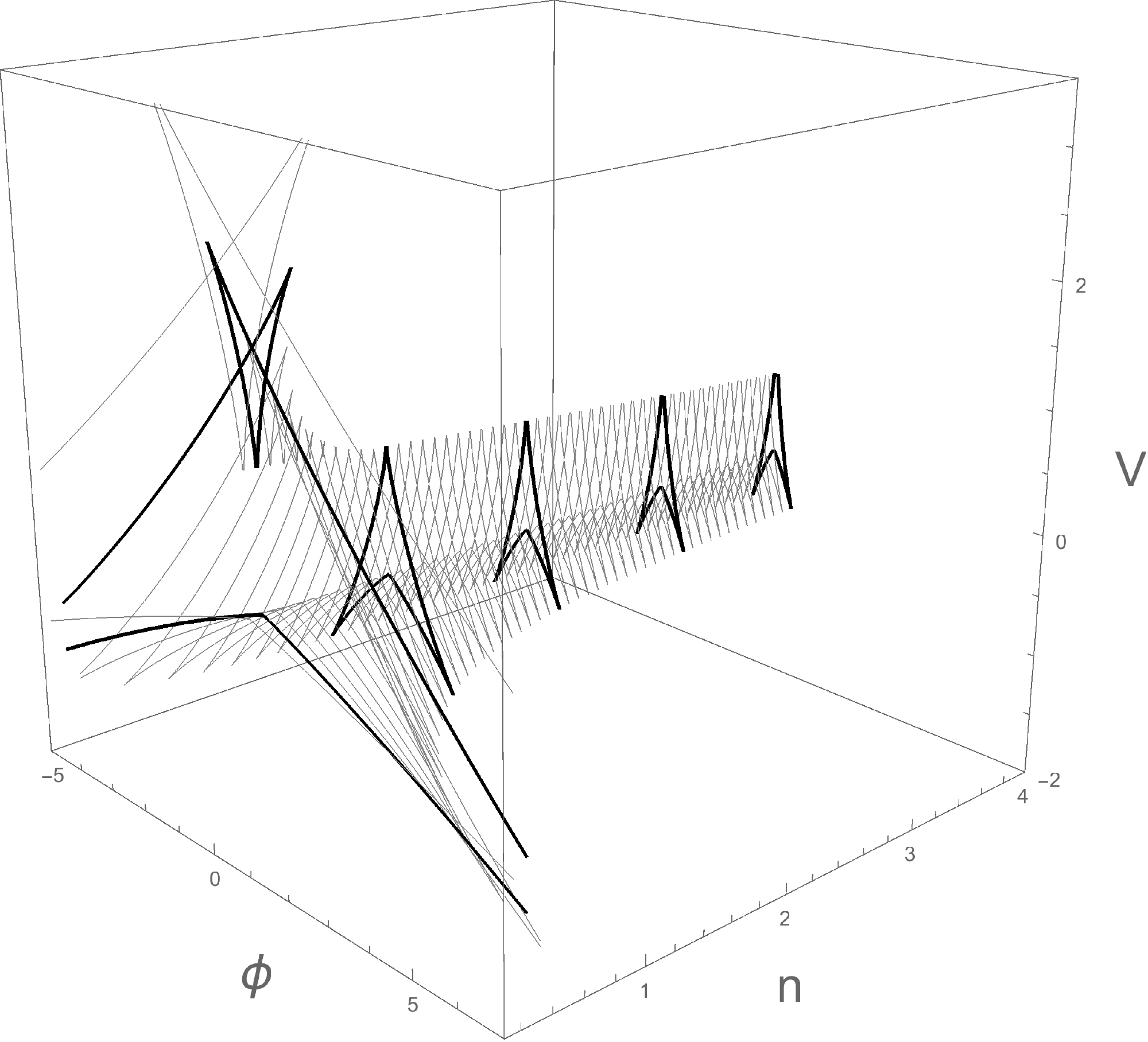}}}\hspace{5pt}
\subfigure[]{
\resizebox*{7cm}{!}{\includegraphics{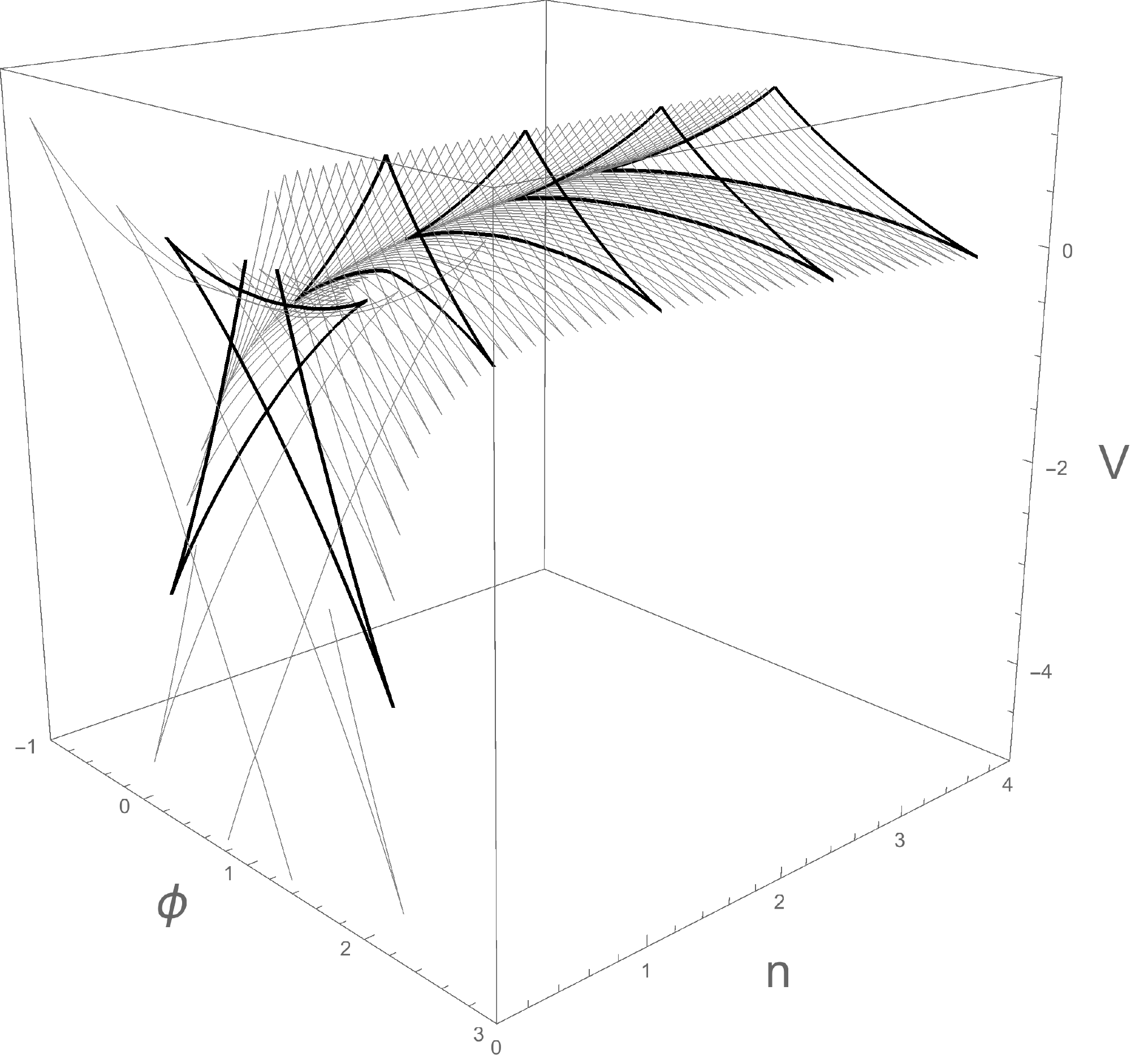}}}
\caption{Scalar potential $V$ as a function of the scalar field $\phi$ and the parameter $n$ for the models studied in this work: Starobinsky (a), Hu-Sawicki (b), complementary exponential (c) and Modified Starobinsky model (d). Black thick lines represents $n=0.25,1,2,3,4$.}\label{scalarpotentials}
\end{figure}

Model (\ref{anotherexp}), proposed in the previous section, meets the necessary conditions to be considered as an extension of the $\Lambda$CDM model, however, performing the expansion to second order at $z R_0$
\begin{multline}\label{expansion1}
   f(R)=z R_0-\mu e^{-n/z^2}\left(1+\frac{1}{z^2}\right) +\left\{R_0-\frac{2 \mu}{z^3}e^{-n/z^2} \left[n \left(\frac{1}{z^2}+1\right)-1\right]\right\}\left(\frac{R}{R_0}-z\right)\\+\frac{\mu}{z^4}\left(-\frac{2 n^2}{z^4}+\frac{(7-2 n) n}{z^2}+3 n-3\right) e^{-n/z^2} \left(\frac{R}{R_0}-z\right)^2+O\left(\frac{R}{R_0}-a\right)^3,
\end{multline}
it is observed at the limit $z\to 0$ that this model does not contain the characteristic $R^2$ term present in Starobinsky expansion (\ref{staroexpansion}), and instead, Eq. (\ref{expansion1}) depends only on R. To avoid this, we will propose the function
\begin{equation}\label{modifiedstarobinsky}
    f(R)=R-\mu  \left[1-\left(1+\frac{R^2}{R_0^2}\right) e^{-\frac{n R^2}{R_0^2}}\right],
\end{equation}
which constitutes a modification of expression (\ref{staro2007}) where the role of the exponent $n$ is played by the exponential function. Note that $R_0=2 \mu  e^{-n} \left(-2 n+e^n-1\right)$ is the constant scalar curvature and $m>0$. Expansion of (\ref{modifiedstarobinsky}) at $z R_0$,
\begin{multline}
    f(R)=z R_0+\mu\left[e^{-n z^2}\left(1+z^2\right)-1\right]+\left[R_0-2 \mu  z e^{-n z^2} \left(n z^2+n-1\right)\right]\left(\frac{R}{R_0}-z\right)\\+e^{-n z^2} \left\{\mu +\mu  n \left(z^2 \left[2 n \left(1+z^2\right)-5\right]-1\right)\right\} \left(\frac{R}{R_0}-z\right)^2+O\left(\frac{R}{R_0}-z\right)^3,
\end{multline}
yields the quadratic term in $R$ when $z\to 0$, that is
\begin{equation}
    f(R)\approx R+\frac{\mu(1-n)}{R_0^2}R^2,
\end{equation}
which has the same functional form of the Starobinsky model of Starobinsky of 1980 [\cite{Starobinsky:1980te}].
This model satisfies (\ref{condition2}) and (\ref{condition1}) in the same way as (\ref{limit}) with ${\mu}\in \mathbb{R}$ and $n>0$. Moreover, from Eq. (\ref{flambda}) it is found
\begin{equation}
    y(R,b)=1-\left[1+\left(\frac{R}{b \Lambda}\right)^2\right] e^{-n (R/b \Lambda )^2},
\end{equation}
with $\Lambda=\mu/2$ and $b=2 R_0/\mu$. The equation between $R$ and $\phi$ is
\begin{equation}\label{modifiedequ}
    R_0^4 (1-\phi ) e^{n R^2/R_0^2}-2 \mu  R \left(n R^2+(n-1) R_0^2\right)=0,
\end{equation}
setting $\mu=R_0=1$, in Fig. \ref{fig:staro3} (d) the solutions of this equation for $n=1, 2, 4, 8$ are shown. Integral (\ref{potentialv}) for this model leads to
\begin{multline}
    v(x)=\frac{e^{-2 n x^2}}{R_0} \left[-2 n^2 x^7+n \left(\frac{3}{2}-4 n\right) x^5+\left(\frac{7}{8}-n (2 n+1)\right) x^3+\frac{(8 n (13-10 n)+21) x}{32 n}\right]\\-e^{-n x^2} \left[2 n \left(x^2+1\right) x \left(x-\frac{2}{R_0}\right)-x \left(x-\frac{4}{R_0}\right)+1\right]-3 \sqrt{2 \pi }\frac{8 (2 n-1) n+7}{128 n^{3/2}R_0}\erf\left(\sqrt{2 n} x\right),
\end{multline}
where $x=R/R0$. Because this model is a modification of Eq. (\ref{staro2007}), the effective potential will have a similar shape of that of Starobinsky, as is depicted in Fig. \ref{fig:severalplots} (h), drawn from those points that solve Eq. (\ref{modifiedequ}), shown in the same figure (g). As a part final of the analysis of this model, the graph of $V$ vs $\phi$ and $n$ is presented in Fig. (\ref{scalarpotentials}).

\subsection{Deviation from GR}
\begin{figure}[h!]
\centering
\subfigure[]{
\resizebox*{8cm}{!}{\includegraphics{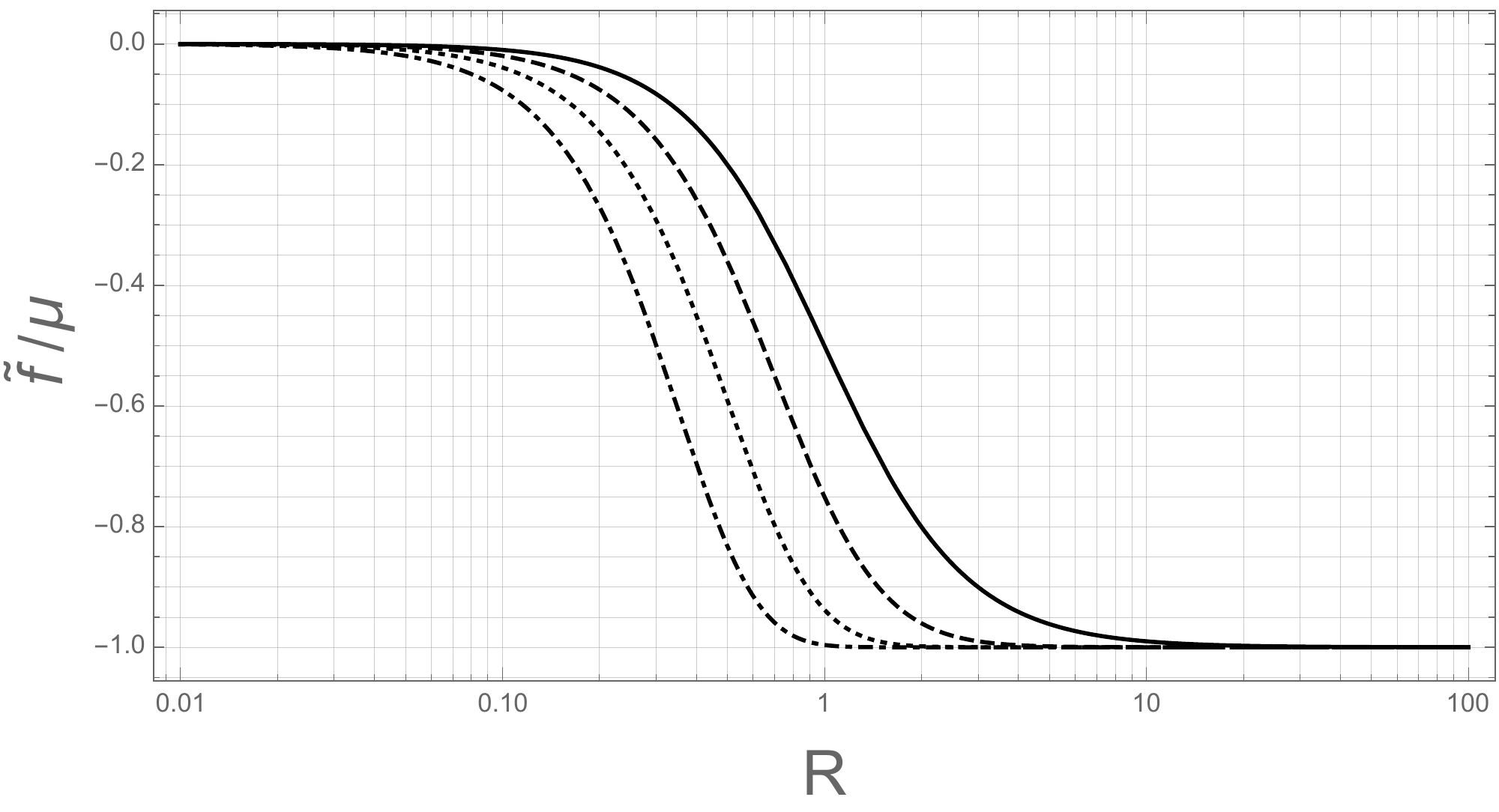}}}\hspace{5pt}
\subfigure[]{
\resizebox*{8cm}{!}{\includegraphics{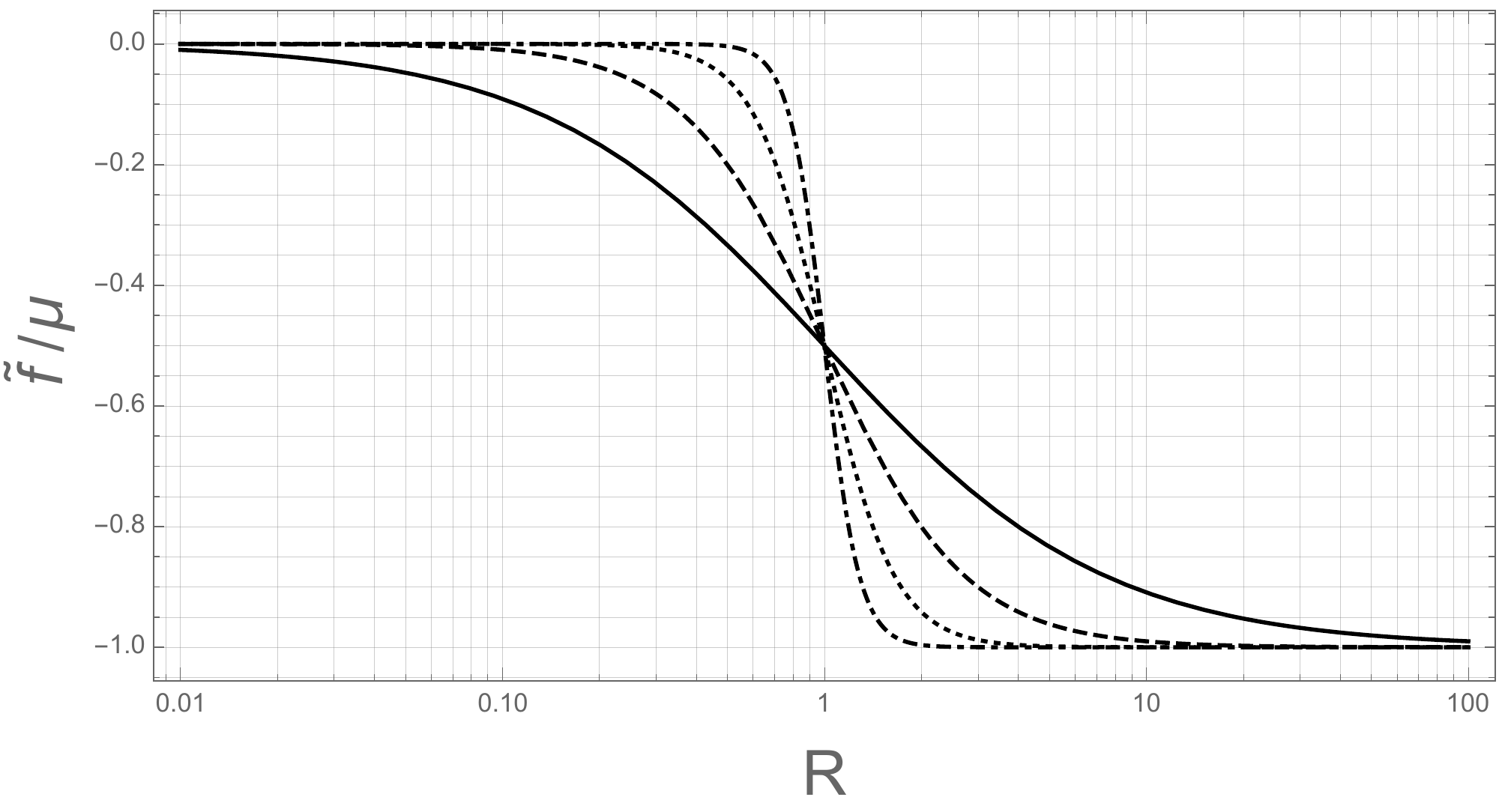}}}
\\
\subfigure[]{
\resizebox*{8cm}{!}{\includegraphics{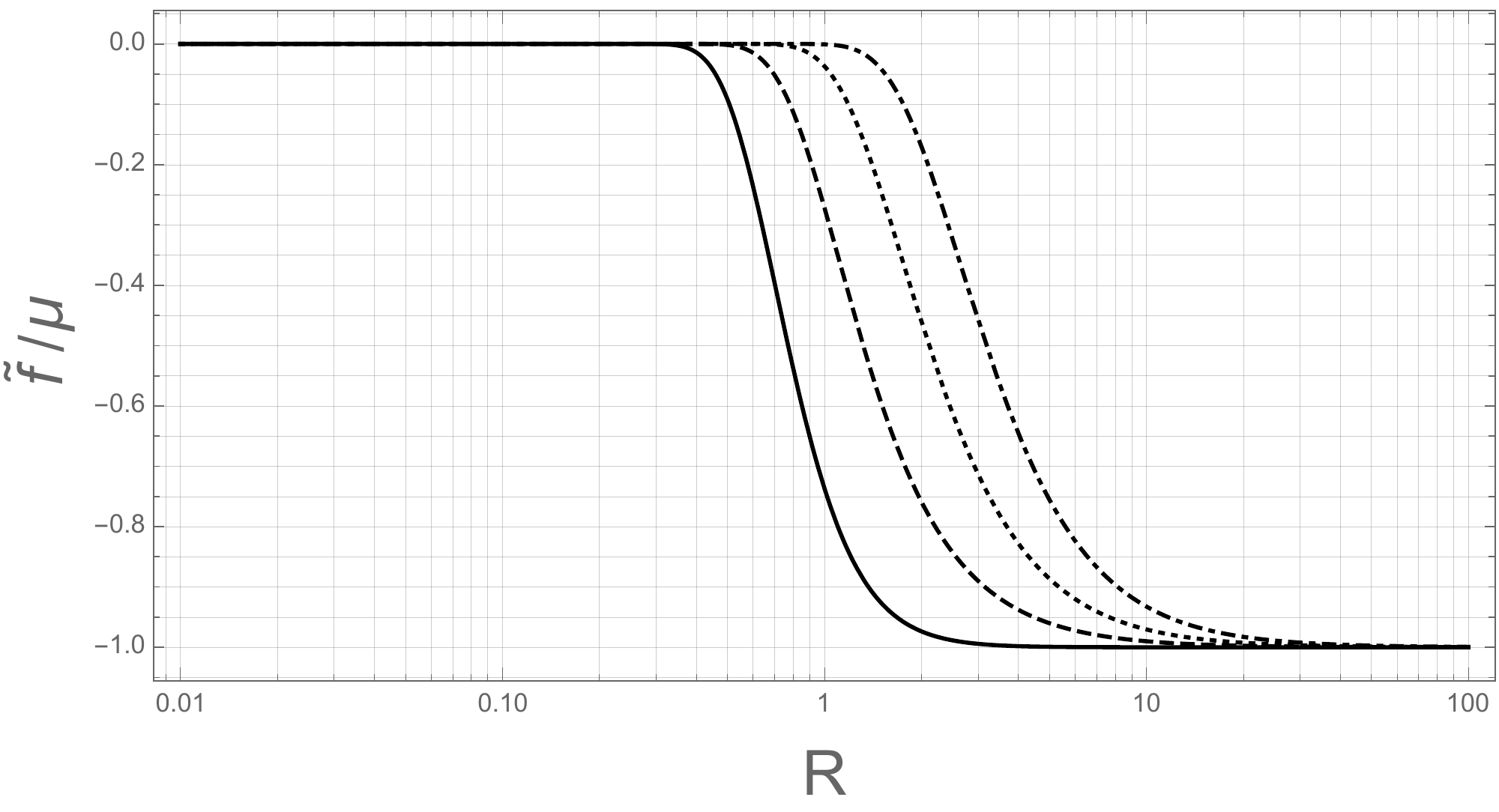}}}\hspace{5pt}
\subfigure[]{
\resizebox*{8cm}{!}{\includegraphics{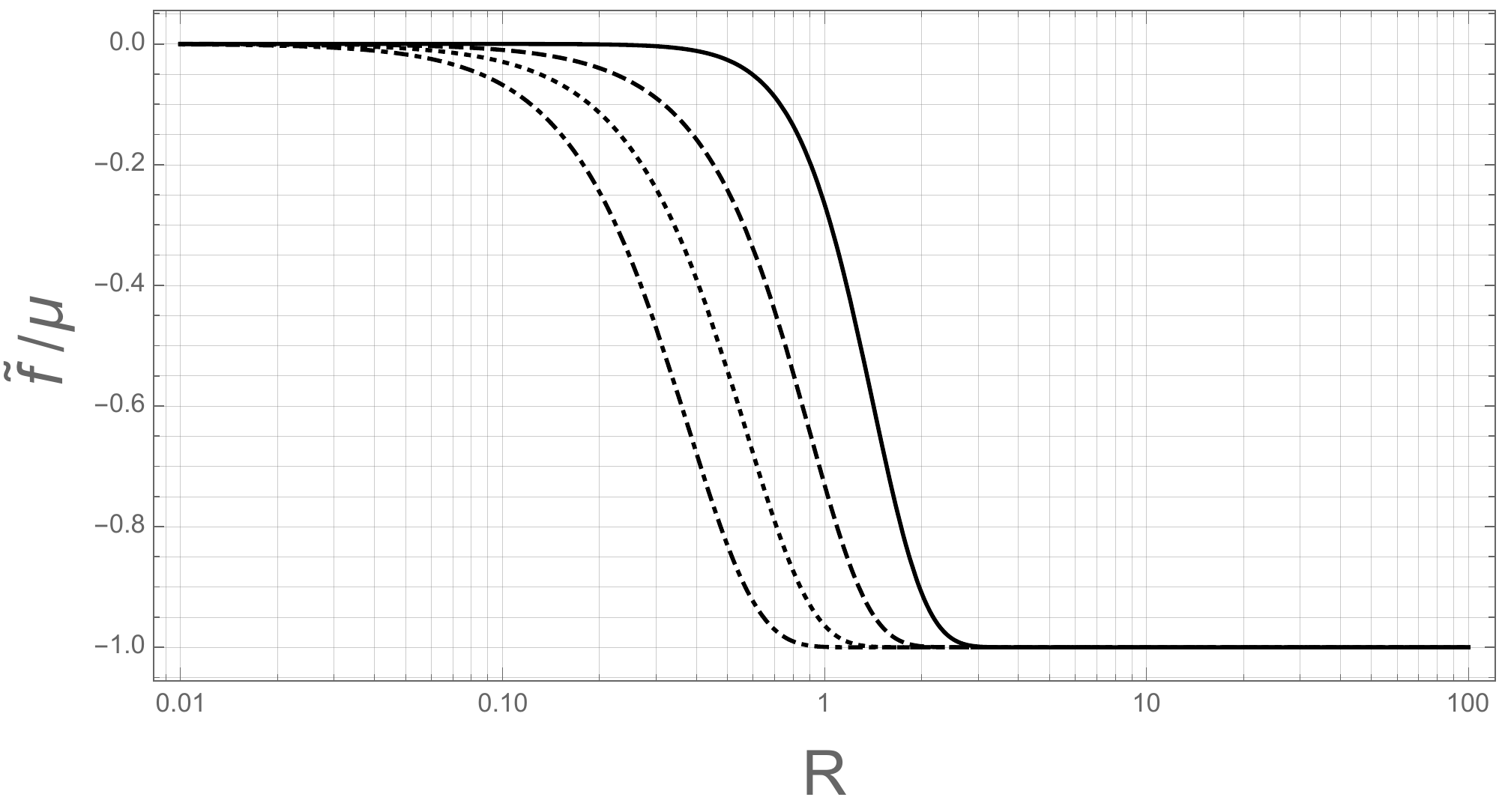}}}
\\
\subfigure[]{
\resizebox*{8cm}{!}{\includegraphics{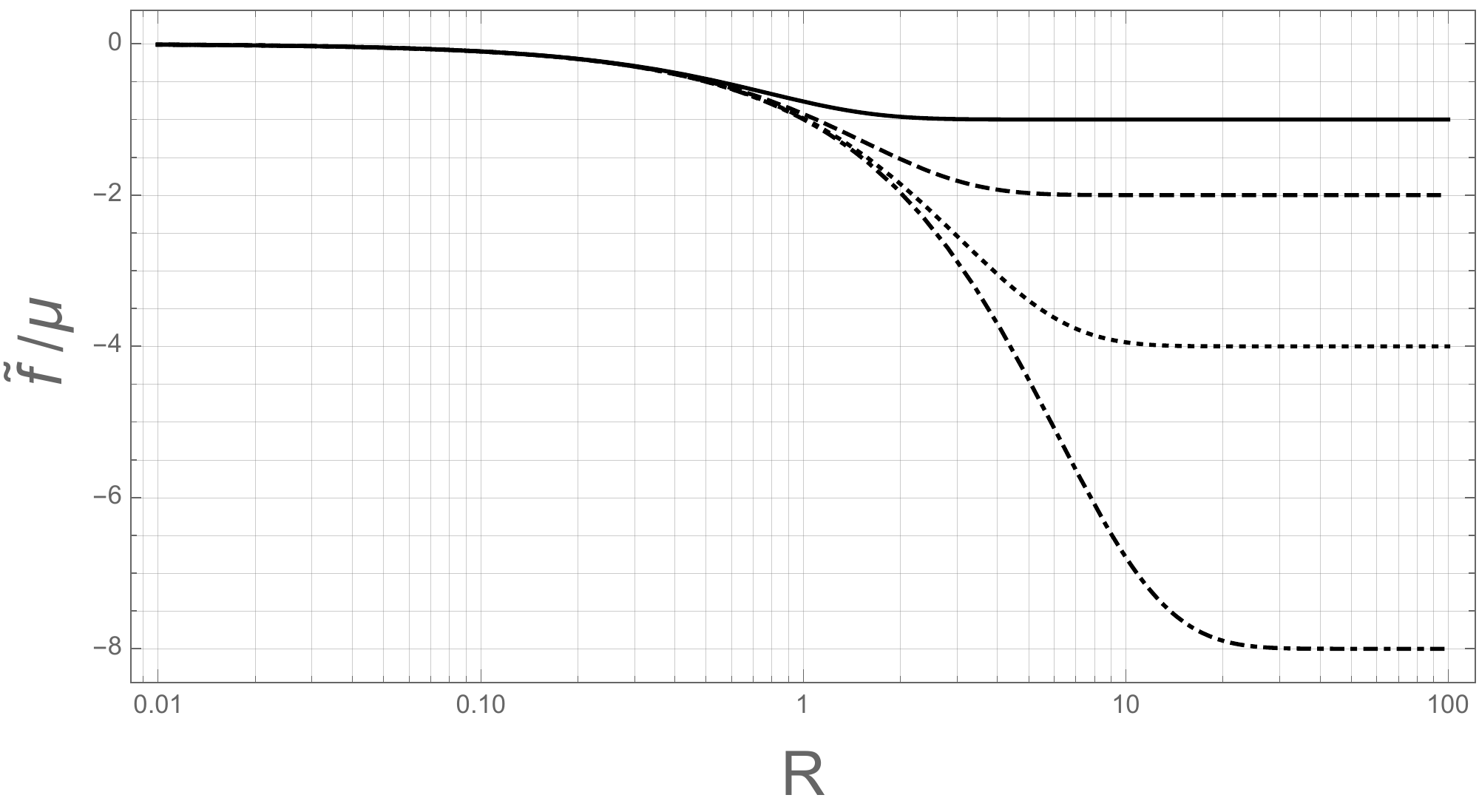}}}\hspace{5pt}
\subfigure[]{
\resizebox*{8cm}{!}{\includegraphics{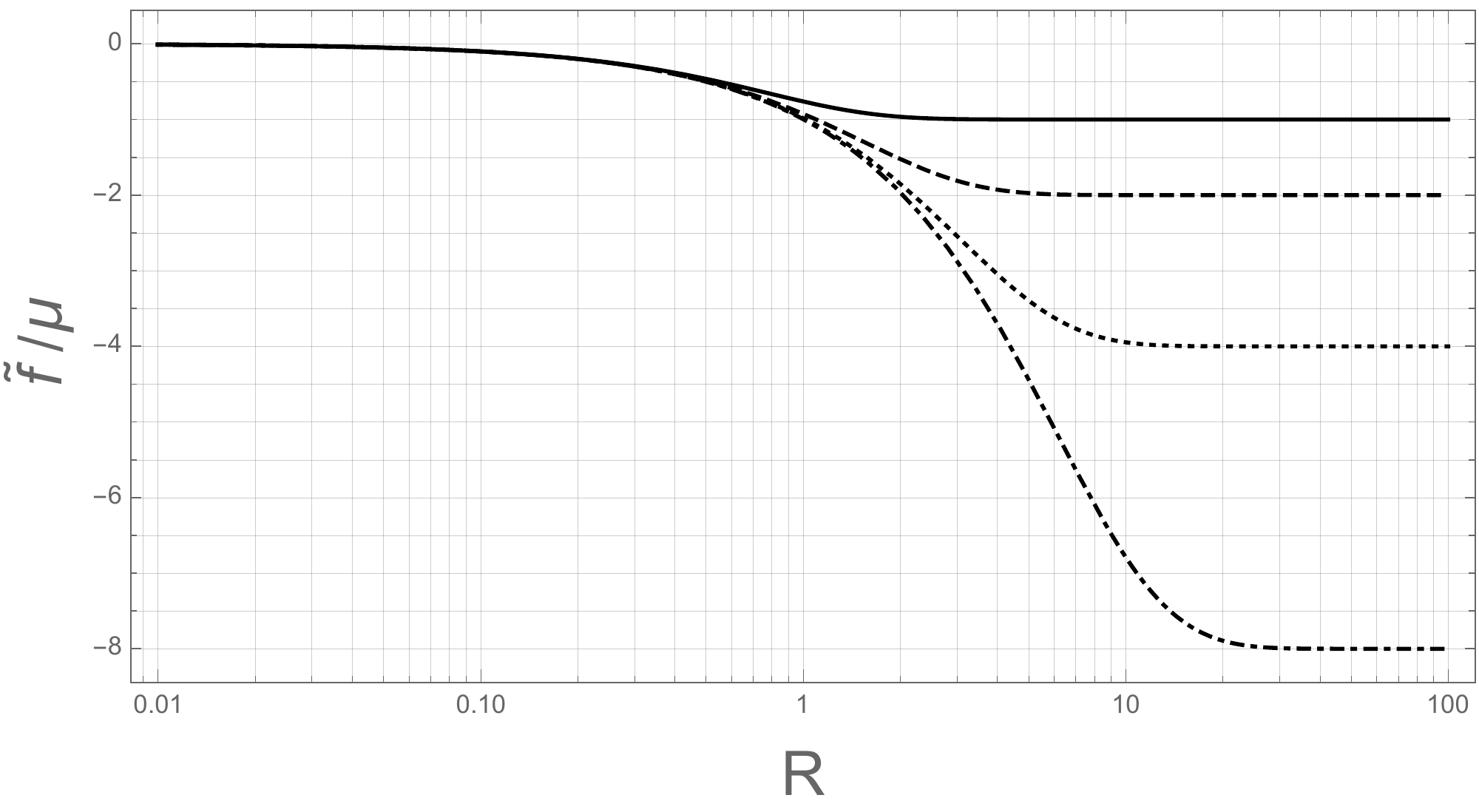}}}
\caption{Behaviour of the net difference between each of the models considered in this work with respecto to GR, measured in the function $\tilde f(R)$. Starobinsky (a), Hu-Sawicki (b), proposed complementary exponential model (\ref{anotherexp}) (c), proposed modified Starobinsky (\ref{modifiedstarobinsky}) (d), Tsujikawa (e) and exponential (f) for $n=R_T=R_E=1, 2, 4, 8$ (black, dashed, dotted and dot-dashed lines). Only in (b) lines intersect at three points: zero, $R=1$ and infinite. For the same $R_T=R_E$ (e) and (f) tend to the same negative limit when $R\to\infty$, however the other models tend to -1, regardless the value of the exponent. }\label{deviation}
\end{figure}

All previous models of $f(R)$ (\ref{staro2007}), (\ref{husawicki}), (\ref{tsujikawa}), (\ref{exponential}), (\ref{anotherexp}) and (\ref{modifiedstarobinsky}) have a parameter $\mu$ that modulates the intensity of deviation of the model with respect to GR, so finally, it is interesting to see the plot of $\tilde{f}(R)/\mu=[f(R)-R]\mu$ for each model to compare them, Fig. \ref{deviation}.

\section{Discussion}\label{sec:discussion}
Modified gravity in the context of an arbitrary function $f(R)$ is indeed a scalar tensor theory with a scalar degree of freedom, and particularly a Brans-Dicke theory with a null parameter $\omega$. In this paper we have made the conformal transformation as general as possible to show the action with the Gibbons-York-Hawking boundary term and the related field equations in the Einstein frame, which are in agreement with the usual field equations in $f(R)$ theory under the inverse transformation. In addition we have defined the effective potential which depends on both geometry and matter-energy, but through trace of the field equations, it turns out to be an integral of a pure geometric term depending only on $f(R)$. With this potential it is possible to find the scalar potential.
Those potentials were calculated, plotted and analyzed for different models recorded in the literature and able to pass the observational test, namely, Starobinsky, Hu-Sawicki, Tsujikawa, Exponential models, and for two new models proposed, which, although look like mathematical toy models, we show that pass the test mentioned above. 
Among all the models analyzed, we delve a little deeper in the Starobinsky model due to its physical relevance, and we show that basically, it consist of a dominant linear and quadratic terms plus 4th-order terms in $R$, and from this express ion it was calculated the scalar field $\psi$, which in turn allowed us to plot the Starobinsky potential in the Einstein frame.
\section*{Acknowledgments}
This work was supported by El Departamento Administrativo de Ciencia, Tecnología e Innovación COLCIENCIAS (Colombia), and by Universidad Nacional de Colombia.

\bibliography{biblio}

\end{document}